\documentclass[jchemphys,10pt,footinbib,letterpaper,onecolumn,superscriptaddress,floatfix,aps]{revtex4-1}

\usepackage[normalem]{ulem}
\usepackage{graphicx}
\usepackage{dcolumn,amsmath}
\usepackage{bm}
\newcommand{\angstrom}{\textup{\AA}}
\usepackage{multirow}

\usepackage[utf8]{inputenc}
\usepackage[T1]{fontenc}
\usepackage{mathptmx}
\usepackage{etoolbox}
\usepackage[colorlinks = true,
        urlcolor  = blue,citecolor=blue]{hyperref}

\makeatletter
\def\@email#1#2{%
 \endgroup
 \patchcmd{\titleblock@produce}
  {\frontmatter@RRAPformat}
  {\frontmatter@RRAPformat{\produce@RRAP{*#1\href{mailto:#2}{#2}}}\frontmatter@RRAPformat}
  {}{}
}%
\makeatother
\usepackage[table]{xcolor}
 
\begin{document}

\preprint{AIP/123-QED}

\title{\Large Revisiting the question of what instantaneous normal modes\\ tell us about liquid dynamics}
\author{Sha Jin}
\affiliation{School of Physics and Astronomy, Shanghai Jiao Tong University, Shanghai 200240, China}
\affiliation{Wilczek Quantum Center, Shanghai Jiao Tong University, Shanghai 200240, China}
\affiliation{Shanghai Research Center for Quantum Sciences, Shanghai 201315, China}
\author{Xue Fan}
\altaffiliation{\color{blue}fanxue2015@shu.edu.cn}
\affiliation{College of Materials, Shanghai Dianji University, Shanghai 201306, China}
\affiliation{Materials Genome Institute, Shanghai University, Shanghai 200444, China}

\author{Matteo Baggioli}
\altaffiliation{\color{blue}b.matteo@sjtu.edu.cn}
\affiliation{School of Physics and Astronomy, Shanghai Jiao Tong University, Shanghai 200240, China}
\affiliation{Wilczek Quantum Center, Shanghai Jiao Tong University, Shanghai 200240, China}
\affiliation{Shanghai Research Center for Quantum Sciences, Shanghai 201315, China}

\date{\today}

\begin{abstract}
The lack of a well-defined equilibrium reference configuration has long hindered a comprehensive atomic-level understanding of liquid dynamics and properties. The Instantaneous Normal Mode (INM) approach, which involves diagonalizing the Hessian matrix of potential energy in instantaneous liquid configurations, has emerged as a promising framework in this direction. However, several conceptual challenges remain, particularly related to the approach's inability to capture anharmonic effects. In this study, we present a set of "experimental facts" through a comprehensive INM analysis of simulated systems, including Ar, Xe, N$_2$, CS$_2$, Ga, and Pb, across a wide temperature range from the solid to gas phase. First, we examine the INM density of states (DOS) and compare it to the DOS obtained from the velocity auto-correlation function. We then analyze the temperature dependence of the fraction of unstable modes and the low-frequency slope of the INM DOS in search of potential universal behaviors. Furthermore, we explore the relationship between INMs and other properties of liquids, including the liquid-like to gas-like dynamical crossover and the momentum gap of collective shear waves. In addition, we investigate the INM spectrum at low temperatures as the system approaches the solid phase, revealing a significant fraction of unstable modes even in crystalline solids. Finally, we confirm the existence of a recently discussed cusp-like singularity in the INM eigenvalue spectrum and uncover its complex temperature-dependent behavior, challenging current theoretical models.

\end{abstract}

\maketitle


\section{Introduction}
Normal modes describe the coherent motion of atoms oscillating at the same frequency and they represent the basis for the study of vibrations in many physical systems from buildings and bridges to molecules and solid state materials \cite{cui2005normal,dykeman2010normal}. The most general dynamics in the linear regime can be indeed reconstructed using a superposition of normal modes that act as building blocks for thermodynamic and transport properties as well. A paradigmatic example is that of harmonic crystals in which the normal mode analysis provides a complete atomic-scale description \cite{moon2023normal}, and leads to the well-established and successful Debye \cite{https://doi.org/10.1002/andp.19123441404} and Einstein \cite{https://doi.org/10.1002/andp.19063270110} models.

An analogous microscopic description of the physical properties of liquids incurs in several difficulties (see \cite{moon2024heat} for an extensive discussion) and remains unachieved \cite{PhysRevLett.89.075508}. Based on molecular flow in liquids, Frenkel interpreted
the Maxwell relaxation time \cite{maxwell1867iv} as a characteristic time between particle jumps and proposed that liquids are solid-like at shorter times (see \cite{trachenko2023theory} for a complete description of these historical developments). Importantly, Frenkel was also the first to discuss
normal modes and their different propagation regimes in liquids (see also important contributions by Zwanzig \cite{PhysRev.156.190}). The hypothesis that normal modes could also exist in liquids, and be a fundamental part of the dynamics, was quickly verified by Rahman, Mandell, and McTague \cite{rahman1976molecular} using a LJ simulated model, giving birth to the concept of ``instantaneous normal modes'' (INMs), that was developed later on to great extent by many authors.\cite{keyes1997instantaneous}.

INMs can be obtained by probing the potential energy surface (PES) at any instant of time and diagonalizing the corresponding instantaneous Hessian matrix that encodes all the information about the local curvature of the PES. In this way, the INM density of states (DOS) can be defined by averaging on several of these instantaneous configurations. Although, at present, this procedure can be easily achieved using computer simulations, what \textit{INM are and are not} remains a challenging question \cite{doi:10.1021/ar00053a001}. In particular, whether the harmonic oscillator picture is just a mere mathematical formal analogy or some kind of coherent oscillatory motion occurs in liquids is a topic of debate. The INM analysis assumes that interactions are harmonic (fully captured by the second derivative of the potential -- Hessian matrix), which is a very questionable assumption in liquids. Nevertheless, this crude approximation might have some validity if the Maxwell picture, where short time dynamics is elastic in nature, is accepted. Importantly, short time dynamics has repeatedly been found to be directly linked to long time dynamics so that the limited applicability of the INM analysis might conceivably be extended to longer time relaxation processes.

Among the various difficulties in this program, it was early realized by Rahman Mandell, and McTague \cite{rahman1976molecular} that, although the INM density of states displays striking similarities with the Fourier transform of the velocity auto-correlation function (VACF), the INM procedure inevitably leads to the presence of negative eigenvalues corresponding to regions in the PES with locally negative curvature. These negative eigenvalues translate into imaginary frequencies, \textit{i.e.}, ``unstable modes''. INMs can be, therefore, divided into stable (positive eigenvalues and real frequencies) and unstable (negative eigenvalues and imaginary frequencies), and the same separation can be applied to the corresponding INM DOS. Let us immediately remark that the ``density of states'' of unstable INM cannot be directly interpreted as a vibrational density of states. More in general, unstable modes cannot be identified with collective atomic vibrations and their physical interpretation appears more subtle and related to the relaxational processes, active in liquids but suppressed in crystals. It is interesting to notice that thermal fluctuations, moving atoms away from equilibrium, can directly induce the emergence of imaginary modes in the harmonic approximation.  As a direct example, let us consider a LJ potential $V=A/r^{12}-B/r^6$: the equilibrium position is $r^*=2^{1/6} (A/B)^{1/6}$ and the second derivative is negative for $r$ larger than about $10\%$ of $r^*$. This gives an imaginary frequency for a finite fraction of atoms that are not performing any diffusing dynamics but rather oscillating around their equilibrium position due to thermal motion.

More in general, the fact that liquids display solid-like oscillatory dynamics at large frequencies, or equivalently short times, it is not a mystery nor a surprise \cite{PhysRevE.51.2654}, and it is experimentally verified, \textit{e.g.}, \cite{doi:10.1073/pnas.1006319107}. These fast dynamics can be well captured by the INM analysis and by the concept of stable INMs, whose density of states is in good agreement with the Fourier transform of the VACF above a certain cutoff frequency (see, for example \cite{doi:10.1021/j100189a029}). Meanwhile, it is well-know that these two quantities are different at low frequencies (compared to an inverse characteristic time $1/\tau$). As a concrete example, the zero frequency value of the INM DOS vanishes while the corresponding limit in the VACF gives the non-zero self-diffusion constant. This is just a manifestation that a harmonic description of liquids cannot apply at long times because of the inevitably anharmonic nature of the potential landscape \cite{10.1063/1.476690} (some anharmonic extensions of the INM approach have been proposed in the past, \textit{e.g.}, \cite{10.1063/1.476768}). It was therefore a surprising discovery \cite{keyes1994unstable,10.1063/1.469947} that INMs, and in particular unstable ones, can still carry important information about the late-time liquid dynamics and that the self-diffusion constant can be reconstructed from the distribution of some of them (even if a concrete characterization of the adjective ``some'' is a challenging task, see \cite{10.1063/1.474822} and related discussions \cite{10.1063/1.475376,PhysRevE.65.026125,10.1063/1.478211,10.1063/1.3701564,10.1063/1.458192}). More recently, it has been proposed that another transport property of liquids -- the shear viscosity -- can be reconstructed from unstable INMs \cite{huang2024microscopic}.

In frequency domain, the INM density of states exhibits interesting features that have been the subject of intense investigation over the past decades. Particular attention has been devoted to understanding the frequency and temperature dependence of the unstable INM DOS \cite{vijay,vijay2,PhysRevE.55.6917,10.1063/5.0158089}, which is commonly represented on the negative frequency axes upon performing a variable transformation $\omega=- \sqrt{|\lambda|}$ with $\lambda$ being the (negative) eigenvalues. Among the various features, the unstable INM DOS displays a universal linear in frequency scaling at low frequencies that is observed in uncountable simulations and derived within several theoretical frameworks \cite{doi:10.1063/1.457564,doi:10.1063/1.465563,doi:10.1063/1.467178,doi:10.1073/pnas.2022303118,doi:10.1073/pnas.2119288119,doi:10.1063/1.479810,PhysRevE.55.6917,doi:10.1063/1.463375}. At present, it is clear that the emergence of this scaling is intimately connected to the existence of unstable modes. In order to avoid any misunderstanding, we stress that the linear scaling of the INM DOS is not in contradiction with numerous experimental results confirming the presence of linearly dispersing acoustic hydrodynamic modes in liquids (\textit{e.g.}, \cite{RevModPhys.77.881,Pilgrim_2006}) that, on the other hand, correspond to a Debye quadratic DOS. In fact, the linear term in the DOS does not arise from propagating modes but rather from overdamped, and quasi-localized, excitations that are abundant in liquids as a consequence of many relaxational mechanisms (see, for example, \cite{brazhkin2024density} for a recent discussion on this point). The quadratic DOS is recovered at low temperatures in the solid state, but only if non-diffusive unstable modes are filtered out (see, for example, \cite{10.1063/1.474968} for a particular method to do that).

Additional interesting properties emerge when the INM DOS is represented as a function of the eigenvalues, rather than frequencies. In particular, it was early observed \cite{sastry2001spectral,taraskin2002disorder}, and recently re-considered \cite{doi:10.1073/pnas.2119288119,Mossa2023}, that the eigenvalue distribution presents a cusp-like singularity at $\lambda=0$ that remains hidden in the frequency representation but carries important information about liquid dynamics as well. We also notice that the properties of INM are strongly related to the topology of the energy landscape. In particular, unstable modes are controlled by the statistics of the saddle points in the potential energy landscape (PEL). The role played by saddles in the supercooled liquid dynamics has been explored in detail in the past (\textit{e.g.}, \cite{PhysRevLett.85.5356,PhysRevLett.85.5360}). These analyses have revealed that the order of the saddles decreases with the temperature and vanishes at the mode coupling temperature $T_{MCT}$, below which the dynamics are mainly localized within the basins. At the same time, the fraction of delocalized unstable modes vanishes at the same temperature that, thus, corresponds to a localization transition for the unstable INM \cite{10.1063/1.3701564,coslovich2019localization}.

The relation (if any) between the INM DOS, the Fourier transform of the VACF (sometimes called the ``density of state function'') and the experimentally measurable density of states of liquids remains an important open question. We notice that a connection between normal modes in liquids and the VACF, and consequently the diffusion constant and the viscosity, was already advanced by Zwanzig in \cite{10.1063/1.446338}. The density of states of liquids at low energies is much less explored experimentally compared to the solid counterpart as standard experimental probes incur in several difficulties. Until very recently, only few studies explored the low-frequency DOS of liquids using inelastic neutron scattering techniques, \textit{i.e.}, Ref. \cite{PhysRevLett.63.2381} for liquid selenium and Ref. \cite{DAWIDOWSKI2000247} for heavy water. More recently, the low-frequency experimental DOS of different liquid systems was investigated in Ref. \cite{doi:10.1021/acs.jpclett.2c00297} and a universal linear scaling was observed in agreement with the expectations from the INM analysis. Furthermore, the temperature dependence of the linear scaling of the experimental DOS was studied in Ref. \cite{jin2024temperature} and compared to the linear scaling of the INM DOS. Both scaling coefficients were found to follow the same exponential dependence on the inverse temperature; nevertheless, the corresponding activation energies were found to be different. Very recently, a liquid-like linear in frequency DOS has been experimentally observed in a solid-state lithium electrolyte at high temperatures \cite{Ding2025}. Moreover, a crossover between a Debye DOS and a linear DOS has been obtained in simulations of amorphous hafnia upon crossing the glass transition temperature \cite{zeng2025thermal}. Despite these encouraging developments, how to define ``a'' liquid density of states, and if such a quantity ultimately exists, remains work in progress.

Arguably, the difficulties in performing a normal mode analysis for liquids, and in identifying a well-defined and unique density of states, are reflected in the challenge of achieving a microscopic description of liquid thermodynamics and collective dynamics \cite{trachenko2015collective}. over the past years, several studies \cite{PhysRevE.104.014103,10.1063/5.0158089,PhysRevResearch.6.013206,PhysRevE.108.014601} have been attempted to use INMs to describe the thermodynamic properties of liquids and in particular their heat capacity. From these analyses, it has emerged that unstable modes cannot be ignored in liquids as considering only stable normal modes would lead to erroneous conclusions. This is somehow consistent with the older two-phase description of liquids, \textit{e.g.} \cite{10.1063/1.1624057}, when the gas-like modes are related to the unstable INM modes as done in Ref. \cite{PhysRevResearch.6.013206}. Interestingly, unstable INM have started to play an important role in glasses and amorphous solids as well \cite{PhysRevE.105.055004,PhysRevLett.127.108003}. For example, unstable INM have been advocated in the past to locate the onset of glass transition \cite{PhysRevLett.74.936}. Finally, it is not yet clear how the INM picture developed for liquids merges into the more traditional normal mode description of solids, for which Debye's law has to be recovered. It is indeed known \cite{Pallikara_2022}, but often discarded, that also crystalline systems at finite temperature possess a non negligible number of unstable modes. It is not completely understood whether these remaining modes are spurious or contain valuable physical information. In fact, one possible interpretation is that these modes are related to local anharmonicities that do not correspond to crossing barriers and moving to a different basin (hence, they do not contribute to diffusion). In other words, these are not necessarily spurious modes but rather indicators of anharmonicity within a potential well. On the other hand, several methods have been developed to filter them out, as for example considering only pure translation modes \cite{10.1063/1.474968}. A final verdict on the nature of these modes is still an open question that goes beyond the scope of this work. 

In summary, there is increasing evidence that instantaneous normal modes might provide a basis for a microscopic description of liquid properties from their atomic-level motion, to their thermodynamics and collective properties. Nevertheless, the question of what we can learn from INM (in particular the unstable ones) is still open, see \textit{e.g.} \cite{doi:10.1063/1.5127821}. In this work, we revisit this question and we provide a comprehensive INM analysis for many liquid systems in order to collect a series of ``experimental facts'' (hopefully) useful to build the long-sought microscopic theory of liquids.

In Section \ref{sec1}, we present our simulation models and the INM analyses carried out. In Section \ref{sec2}, we take Argon as a case study and we perform an extensive analysis using several methods to explore the role of INM in such a system. We then extend the analysis to a large class of liquids including Xe, N$_2$, CS$_2$, Ga and Pb in order to verify the universality of our findings. In Section \ref{sec4}, we select a set of topics that are part of the broader question of what we can learn from INM about liquids and we provide several ``experimental facts'' within this debate. Finally, in Section \ref{sec5}, we conclude and discuss the most interesting open directions following from our results.

\section{Methods and models}\label{sec1}
We start by describing in full detail the simulations models considered in our study and the methods used. 
\subsection{Interaction potential}
In this work, all simulations are conducted via the Large-scale Atomic/Molecular Massively Parallel Simulator (LAMMPS) \cite{lammps}. The computer simulation of metals was conducted using the pair interparticle interaction method, along with the incorporation of a more advanced embedded atom model (EAM) potential technique. In the EAM, an additional form of collective interaction is introduced alongside the conventional pair interparticle potentials. The expression for the potential energy of a metal is given as
\begin{equation}\label{EAM-potential}
U =\sum_{i}\Phi(\rho_{i}) + \sum_{i<j}{\phi(r_{ij})},
\end{equation}
where $\Phi$$(\rho_{i})$ is the potential of embedding the $i$th atom depending on the "effective electron density" $\rho_i$ at the location of the atom's center and it is determined by the surrounding atoms
\begin{equation}\label{pair-potential}
\rho_{i} =\sum_{j}\Psi(r_{ij}),
\end{equation}
where $\Psi$$(r_{ij})$ is the contribution to electron density of neighbor atom $j$. The second term in Eq.~\eqref{EAM-potential} represents the usual pair potential. 
In this work, the interactions between metal crystals of Ga \cite{Belashchenko} and Pd \cite{Wang_2019} atoms are described with EAM potentials independently, which are widely applied in respective metallic systems \cite{Mokshin2015, Chtchelkatchev2016, Belashchenko2017, 10.1063/5.0031185}. The rest of non-metal systems of particles interact via the 12-6 Lennard-Jones (LJ) potential,
\begin{equation}\label{LJ-potential}
\phi_{\alpha\beta}(r_{ij})=4\epsilon_{\alpha\beta}\left[\left(\frac{\sigma_{\alpha\beta}}{r_{ij}}\right)^{12}-\left(\frac{\sigma_{\alpha\beta}}{r_{ij}}\right)^{6}\right],
\end{equation}
where $i$ and $j$ stand for $i$th and $j$th atom, $\alpha$ and $\beta$  for the type of atoms, and $\epsilon_\alpha$$_\beta$  and $\sigma_\alpha$$_\beta$ are the LJ parameters. All parameters chosen to approximate the intermolecular potential for each system are list in Table \ref{tab1}. For nitrogen molecules, the intramolecular interaction is approximated by the harmonic type potential as
\begin{equation}\label{NN-potential}
U_{NN}(r)=\frac{1}{2}K_{NN}(r-r_e),
\end{equation}
where $K_{NN}$ is the force constant which is measure of interatomic bond between nitrogen atoms in N$_2$ molecule, $r$ is the bond length, and $r_e$ is the bond length at equilibrium. We used a bond spring constant $K_{NN}$ = 13800 KJ/(mol) and a bond length $r_e = 1.0975$. At the same time, the harmonic type potential for CS$_2$ molecules \cite{doi:10.1080/00268978800100851} is given by
\begin{equation}\label{CS2-potential}
U_{intra}=\sum_{bonds}\left[K_1\{\left(r_1-r_0\right)^2+\left(r_2-r_0\right)^2\}+K_2\left(r_1-r_0\right)\left(r_2-r_0\right)\right]+\sum_{angles}K_3\left(\theta-\theta_0^2\right)^2,
\end{equation}
where $r_1$ and $r_2$ are the bond lengths of C-S, and $\theta$ is the bond angle of S-C-S in the CS$_2$ molecule. The parameters used in equation \eqref{CS2-potential} are shown in Table \ref{tab2}.

\begin{table*}[h]
\caption{Lennard-Jones energy and length parameters.}
\label{tab1}
\begin{ruledtabular}
\renewcommand\arraystretch{1.5}
\begin{tabular}{ccccl}

 Molecule&Pair&$\sigma (\angstrom)$&$\epsilon$(Kcal/mol)&$d_{\text{cutoff}}$ $(\angstrom)$
 \\ \hline

Ar \cite{doi:10.1139/v77-418} & ArAr &3.465&0.225 &8\\
Xe \cite{doi:10.1139/v77-418} &XeXe &4.05&0.457 & 10\\
   
     N$_2$ \cite{doi:10.1080/00268979000101011} &NN&3.31&0.0715 &12\\
  
      \multirow{3}{*}{CS$_2$ \cite{doi:10.1080/00268978100100861}}& CC &3.35&0.1017 &\\
  
     \multicolumn{1}{c}{}&SS &3.52&0.3637 &12\\
      
      \multicolumn{1}{c}{}& CS &3.44&0.1923 &\\
\end{tabular}
\end{ruledtabular}
\end{table*}

\begin{table*}[h]
\caption{The intramolecular potential parameters for the CS$_2$ model.}
\label{tab2}
\begin{ruledtabular}
\renewcommand\arraystretch{1.5}
\setlength{\tabcolsep}{0pt} 
\Large
\resizebox{0.6\linewidth}{!}{
\begin{tabular}{cc}

 Parameters&Values
\\ \hline

$K_1$ & 2250 KJ/(mol$\cdot \angstrom^2$)\\
     
      $K_2$ & 360 KJ/(mol$\cdot \angstrom^2$)\\
   
     $K_3$ & 169 KJ/(mol$\cdot \angstrom^2$)\\

     $r_0$ &1.55 \angstrom \\
      
     $\theta_0$ & 180$^{\circ}$\\
\end{tabular}
}
\end{ruledtabular}
\end{table*}

\subsection{Simulation details}

\begin{figure}[ht]
    \centering
    \includegraphics[width=0.75\linewidth]{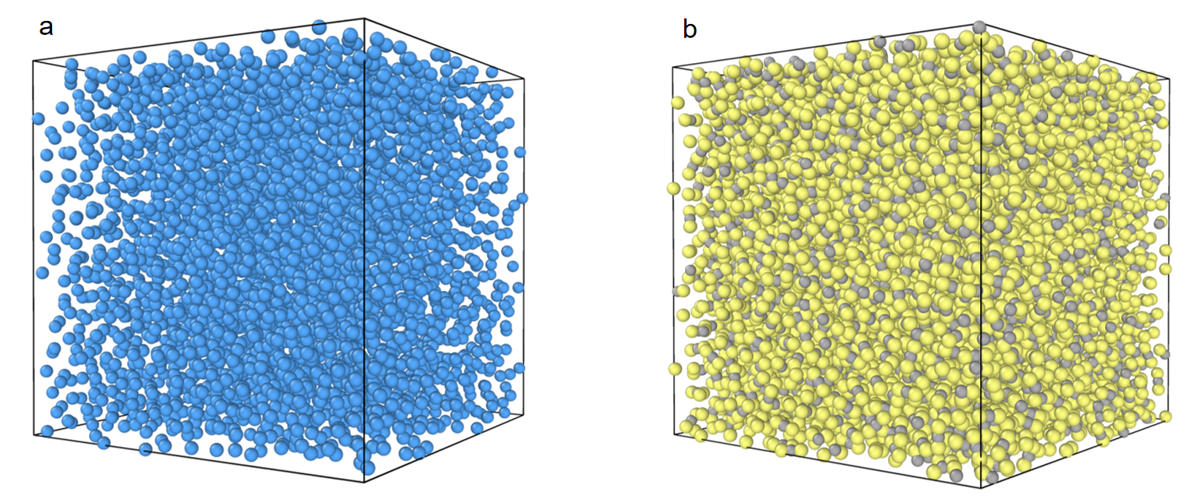}
    \caption{MD simulation systems for liquid Ar and CS$_2$. The snapshot of liquid Ar \textbf{(a)} at $T = 100$ K, $P = 40$ bar and liquid CS$_2$ \textbf{(b)} at $T = 300$ K, $P = 1.01 $ bar. Figures made with OVITO software \cite{Stukowski_2010}.}
    \label{fig0}
\end{figure}

For the four monatomic systems, a Face-Centered Cubic (FCC) lattice structure of 4000 atoms with lattice constant of $a_A$$_r$=$5.268\,\angstrom$ \cite{PhysRev.111.1470}, $a_X$$_e$= $6.347\,\angstrom$ \cite{PhysRevB.24.4753}, $a_G$$_a$= $4.4\,\angstrom$ \cite{Belashchenko} and $a_P$$_b$= $4.95\,\angstrom$ \cite{Wang_2019} was respectively used as an initial input structure for the MD simulations. For N$_2$ and CS$_2$ systems, the MD simulations were performed with 500 N$_2$ and 3375 CS$_2$ molecules in either one cubic box of dimension 30.0 $\angstrom$  and 80.0 $\angstrom$ respectively. All samples were simulated with periodic boundary conditions (PBC).
In the process of samples preparation, after energy minimization was performed using steepest descent algorithm to bring the initial configuration in minimum potential, all samples relaxed at least $5$ ns (MD time step is $1.0$ fs)  to reach their respective equilibrium at a series of temperatures and specific pressures based on phase diagram to keep them liquid. In this process, the isobaric and isothermal (NPT) ensemble was used with the sample size being adjusted to give target pressure. After that, the canonical (NVT) ensemble MD was conducted at different temperatures  for data collections and analysis with an interval of $100$ fs during a $1$ ns simulation. The MD simulation systems of Ar and CS$_2$ are shown in Figure \ref{fig0}.

The pressures of all of our simulated systems are provided in the Table \ref{tab3}. In our simulation systems, the temperature range for argon is the most extensive, covering all three phases: solid, liquid, and gas. The approximate melting and boiling points for argon are provided in Table \ref{tab3}. For xenon, plumbum, and gallium, the simulated temperature range includes both solid and liquid phases, with the melting temperature presented in Table \ref{tab3} as well. It is important to highlight that, since the phase transition dynamics are not the focus of our work, our simulations were performed at fixed temperature rather than simulating the phase transition process itself. The melting point was determined based on the system’s state at different temperatures rather than through direct observation of the transition dynamics (see details in Appendix \ref{ioio}). As a result, the identified melting points are not accurate and may slightly differ from the experimental measurements. To provide context, experimental data from the literature are also included in the table for comparison. For N$_2$ and CS$_2$ systems, we focused on simulating the liquid phase; both the melting and boiling points for these two systems presented in Table \ref{tab3} are taken from experimental data.

\begin{table*}[h]
\caption{Melting and boiling temperatures for the systems considered in this work.}
\label{tab3}
\begin{ruledtabular}
\renewcommand\arraystretch{1.5}
\begin{tabular}{cccc}

 System&Pressure (bar)&Melting temperature (K)&Boiling temperature (K)
 \\ \hline

Ar  & 40 &92&130\\
Xe  &1.01 &145,  161 \cite{callister2020materials}& 165 
 \cite{callister2020materials} \\
Pb &1.01 &740, 600 \cite{wohlfarth2001crc}&2013 
 \cite{wohlfarth2001crc} \\
Ga &1.01 &320, 303 \cite{2005NIST}&2676 \cite{zhang2011corrected}\\
N$_2$&70.92&63 \cite{Engineeringtoolbox_2024}&127 \cite{2005NIST}\\
CS$_2$&1.01 &161 \cite{2005NIST}&319 \cite{2005NIST} \\
 
\end{tabular}
\end{ruledtabular}
\end{table*}

\subsection{Definition of the physical observables}
In this Section, precise definitions of the key physical quantities considered in this work are provided.

The velocity auto-correlation function (VACF), $Z(t)$, is an essential time-correlation function in physics and chemistry that provides extensive information about molecular-structural and hydrodynamic properties of matter \cite{hansen2013theory}. The Fourier transform of $Z(t)$ is defined as the density of state function, $S(\omega)$. More precisely,
\begin{equation}\label{vacfdef}
    Z(t)=\frac{\langle \vec{v}(t) \vec{v}(0) \rangle}{\langle \vec{v}(0) 
 \vec{v}(0) \rangle }\,,\qquad S(\omega)=\int_{-\infty}^{\infty} e^{-i \omega t} Z(t) dt.
\end{equation}
On the other hand, the instantaneous normal mode analysis provides an instantaneous picture of liquid dynamics that has been widely used in the past \cite{keyes1997instantaneous}. The INM method involves calculating the normal modes of a system at a given instantaneous configuration, providing insight into the vibrational properties and stability of the system at specific moments in time. It starts from the dynamical matrix that characterizes the curvatures of the local potential energy surface. The dynamical matrix is a $3N\times 3N$ matrix whose elements are defined as follows:
\begin{equation}
H_{i\mu,j\nu}(\mathbf{R})=\frac{1}{\sqrt{m_i m_j}}\frac{\partial^2V}{\partial r_{i,\mu}\partial r_{j,\nu}}.
\end{equation}
Here, $i,j=1,...,N, \mu,\nu=x,y,z$. $N$ is the number of total atoms in the system. $\mathbf{R}\equiv {\mathbf{r}_1,...,\mathbf{r}_N}$ represents each liquid configuration and $\mathbf{r}_i$ is the position of the $i$th atom. V is the potential energy and $r_{i,\mu}$ represents the $\mu$-coordinate of the $i$th atom. 
The instantaneous normal mode frequencies $\omega_i$ are defined as the square roots of the eigenvalues of the dynamical matrix $\lambda_i$. The INM density of states, $\rho(\omega)$, is then obtained by averaging over different instantaneous configurations,
\begin{equation}\label{inmsdef}
\rho(\omega)=\left<\frac{1}{3N}\sum_i^{3N}\delta(\omega _i-\omega)\right>.
\end{equation}
Due to locally negative curvature regions in the potential energy landscape, INMs can be divided into stable modes (positive eigenvalues) and unstable modes (negative eigenvalues). We define the fraction of unstable modes, $f_u$, as follows:
\begin{equation}\label{fudef}
f_u=\frac{N_u}{3N},
\end{equation}
where $N_u$ presents the number of unstable modes. The complementary fraction of stable mode $f_s$ is just its complement, \textit{i.e.} $f_s=1-f_u$. In an ideal zero temperature crystal, one expects $f_u=0$ and $f_s=1$. 

Finally, we define the slope of the INM DOS as $a(T)$ from the low-frequency behavior,
\begin{equation}\label{atdef}
    \rho(\omega)=a(T) |\omega|+\dots
\end{equation}
where negative values of $\omega$ correspond to imaginary frequencies, and the ellipsis stands for higher order corrections in $\omega$.

From the MD simulation, we also calculate the diffusion coefficient, $D$, and the pair correlation function, $g(r)$, to gain a comprehensive understanding of the dynamic and structural properties of the simulated systems. The diffusion coefficient, $D$, quantifies the rate of particle diffusion and is determined using the Einstein relation, which relates $D$ to the particle mean square displacement (MSD),
\begin{equation}\label{Ddef}
    D=\lim_{t \to \infty}\frac{1}{6t} \left<\lvert \mathbf{r}(t)-\mathbf{r}(0)\rvert^2\right>,
\end{equation}
where $\mathbf{r}(t)$ represents the position of a particle at time $t$, $\mathbf{r}(0)$ is its initial position, and $\langle \cdot \rangle$ denotes an ensemble average.
The pair correlation function, $g(r)$, provides insight into the structural arrangement of particles within the system. It is defined as:
\begin{equation}\label{grdef}
    g(r)=(V/N)\left[n(r)/4\pi r^2\Delta r\right]
\end{equation}
Here, $V$ represents the volume of the system, $N$ is the total number of particles, and $n(r)$ is the number of atoms that are situated at a distance between $r$ and $r+\Delta r$ from a given particle. The term $4\pi r^2\Delta r$ is the volume of the spherical shell at distance r with thickness $\Delta r$. Physically, $g(r)$ reveals how the probability of finding a particle at a distance $r$ from a reference particle deviates from the ideal gas limit. Precisely, $g(r)= 1$ indicates that the particles are distributed randomly at distance $r$, while $g(r)$ greater than $1$ suggests the presence of local structural order. Conversely, $g(r)$ less than $1$ indicates a depletion of particles at that distance, suggesting the presence of repulsive interactions or an exclusion zone around each particle.

\section{Argon as a benchmark case}\label{sec2}
\begin{figure}[h]
    \centering
    \includegraphics[width=0.45\linewidth]{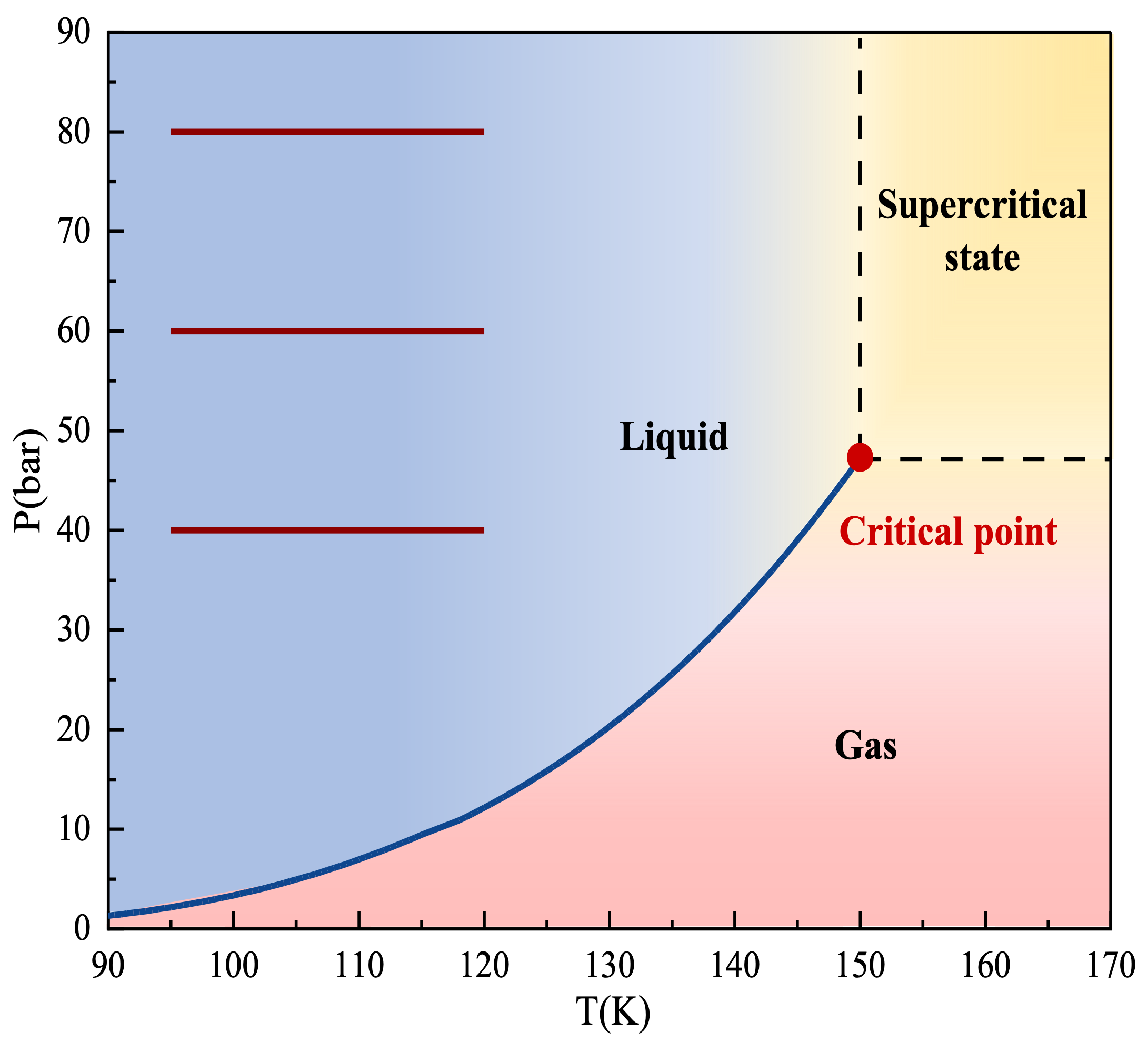}
    \caption{A zoom of the phase diagram of Argon (Ar) indicating the liquid and gas phases as well as the critical point at approximately $151$ K and $48.55$ bar. The horizontal red lines indicate the set of thermodynamic conditions explored in this manuscript. The blue dots correspond to the dynamical crossover points discussed below.}
    \label{fig:1}
\end{figure}

We take the case of Argon (Ar) as a paradigmatic simple example to study liquid properties within the INM framework. The phase diagram of Argon in the regime of interest is shown in Fig.\ref{fig:1}. The black line indicates the line of first-order phase transitions separating the liquid (light blue) and the gas (light green) phases. The critical point, indicated with a red circle, is located at $T_c\approx 151$ K and $P_c \approx 48.55$ bar \cite{1967Critical}. The data for the phase diagram are collected from the National Institute of Standards and Technology (NIST) database \cite{2005NIST}. The pink region is the supercritical phase beyond the critical point. The horizontal red isobar lines indicate some of the regions that have been numerically investigated in this work. Other investigated isobar lines with much higher pressure are not displayed in Fig.\ref{fig:1}.

Because of space limitations, in the main text we focus only on the isobar line at $P=40$ bar and we explore a wide region of temperatures from $95$ K to $160$ K. This isobar line crosses the first order phase transition between liquid and gas at approximately $T \approx 146$ K. 

\begin{figure}
    \centering
    \includegraphics[width=0.8\linewidth]{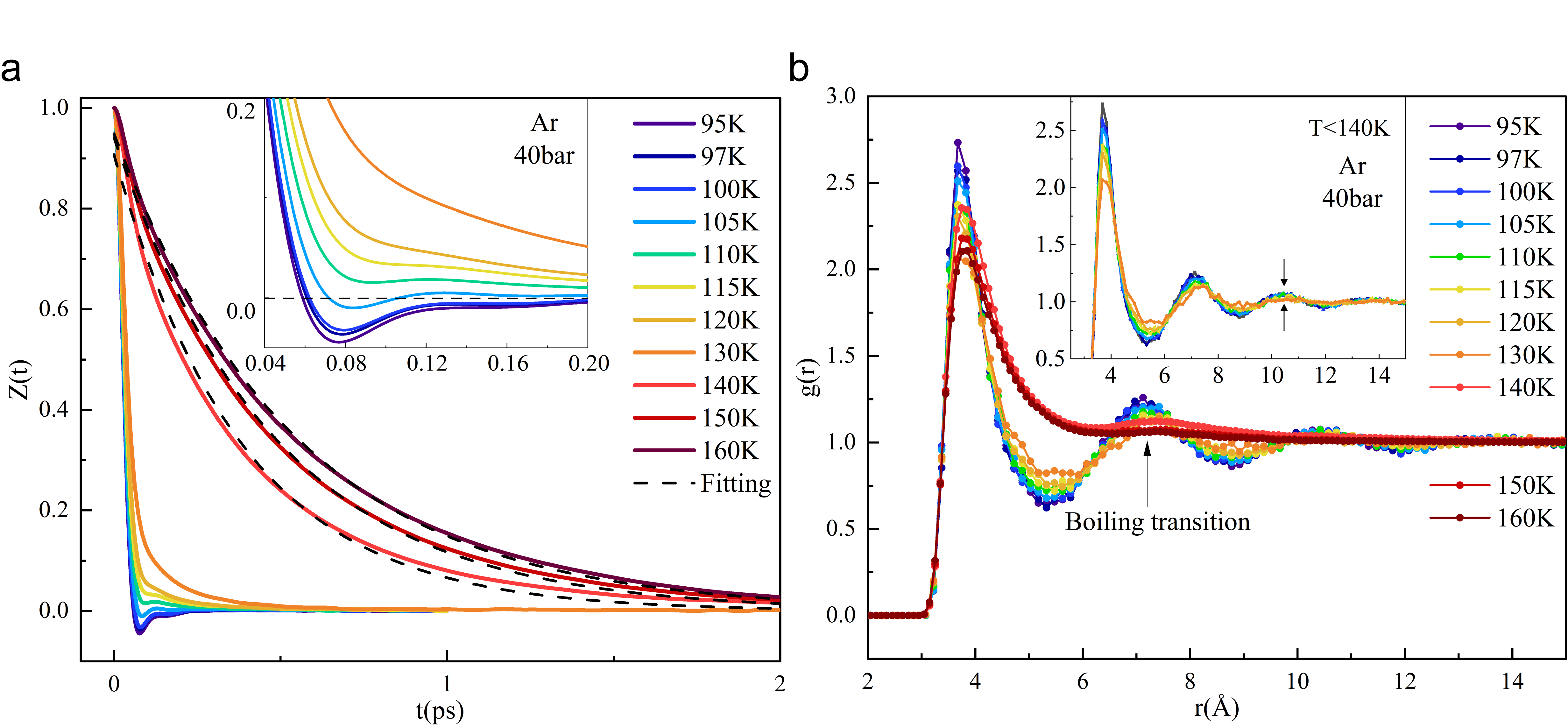}
    \caption{\textbf{(a)} The velocity auto-correlation function (VACF) $Z(t)$ defined in Eq.\eqref{vacfdef} for Argon at $P=40$ bar and different values of temperature. The inset zooms on the temperature values below $T=140$ K to emphasize the dynamical crossover (Frenkel line) that happens at $T\approx 115$ K. The dashed black lines are fits to a purely exponential function as discussed in the main text. \textbf{(b)} The corresponding pair correlation function $g(r)$. The arrow indicates the structural signatures in the second peak of $g(r)$ of the first order gas-liquid phase transition that happens between $130$ K and $140$ K. The inset shows the disappearance of intermediate range order (third peak in $g(r)$), whose onset is consistent with the dynamical crossover in the inset of panel (a).}
    \label{fig:2}
\end{figure}

We start our analysis by considering the behavior of the velocity auto-correlation function (VACF) $Z(t)$ and the pair distribution function (PDF) $g(r)$ that are shown in panels a and b of Fig.\ref{fig:2} respectively. At high temperature, deep in the gas regime, $Z(t)$  decays monotonically and its form is well approximated by an exponential function $\exp(-\gamma t)$ (shown with black dashed line), as predicted by kinetic theory under the assumption of pure Brownian motion \cite{hansen2013theory}. Upon decreasing the temperature, $Z(t)$ remains a monotonically decreasing function but it is not anymore well approximated by a pure exponential function. This indicates the onset of correlations and the tendency towards dense liquid dynamics. A big jump is observed between $140$ K and $130$ K that corresponds approximately to the boiling temperature of Argon at this pressure. Even in the liquid phase, at least up to a certain ``critical'' temperature, the velocity auto-correlation function remains a monotonically decreasing function.

More importantly, as emphasized in the inset of panel a in Fig.\ref{fig:2}, the velocity-autocorrelation function displays a drastic change of behavior around $115$ K where it starts developing a minimum that transforms into marked oscillations by decreasing the temperature further. This drastic change of behavior corresponds to a transition between gas-like dynamics to liquid-like dynamics and it is known in the literature as the Frenkel line \cite{10.1063/PT.3.1796,PhysRevE.85.031203,PhysRevLett.111.145901}. The Frenkel line has been studied in several liquids in the supercritical phase \cite{doi:10.1021/acs.jpclett.8b01955,Bolmatov2013,doi:10.1021/jz5012127,PhysRevB.95.134114,Bolmatov2015,Fomin_2018,proctor2019transition,fomin2015dynamical,doi:10.1021/jp2039898,COCKRELL20211}, but also in complex dusty plasmas \cite{PhysRevResearch.5.013149,PhysRevE.107.055211} and active granular fluids \cite{jiang2025experimental}. 

In order to investigate structural properties, in panel b of Fig.\ref{fig:2} we have computed the pair distribution function $g(r)$ in the same range of temperatures and constant pressure $P=40$ bar. At low temperatures, we observe the typical behavior of a dense liquid with a first strong peak around $4\,\angstrom$ and other weaker peaks around $7\,\angstrom$, $10.5\,\angstrom$ and so on. By increasing the distance $r$, the peaks disappear signaling the absence of long-range order, as one might expect in a liquid. As shown in the inset of panel b of Fig.\ref{fig:2}, by increasing the temperature but still below the boiling temperature of $\approx 140$ K, we observe that the higher peaks (above the first two) corresponding to intermediate-range order disappear. For example, as indicated by the black arrows, the intensity of the third peak located at approximately $r \approx 10.5 \,\angstrom$ vanishes around $115-120$ K. Interestingly, these temperatures are extremely close to the location at which oscillations in the VACF $Z(t)$ disappear (panel a in Fig.\ref{fig:2}). As already proposed in \cite{10.1063/1.4844135}, this suggests the existence of a structural change in the intermediate-range order that might be associated to the Frenkel line.

By increasing further the temperature, across the boiling transition, the structure of the system changes drastically. First, all the peaks apart from the first one disappear, implying that intermediate-range order is totally lost. Second, the minimum around $r \approx 5 \,\angstrom$ disappears as well and $g(r)$ displays a sharp jump at that length scale. Finally, despite the height of the first peak is reduced, it does not vanish. This implies that we are still far from the totally uncorrelated ideal hard-sphere gas regime.

\begin{figure}
    \centering
    \includegraphics[width=0.8\linewidth]{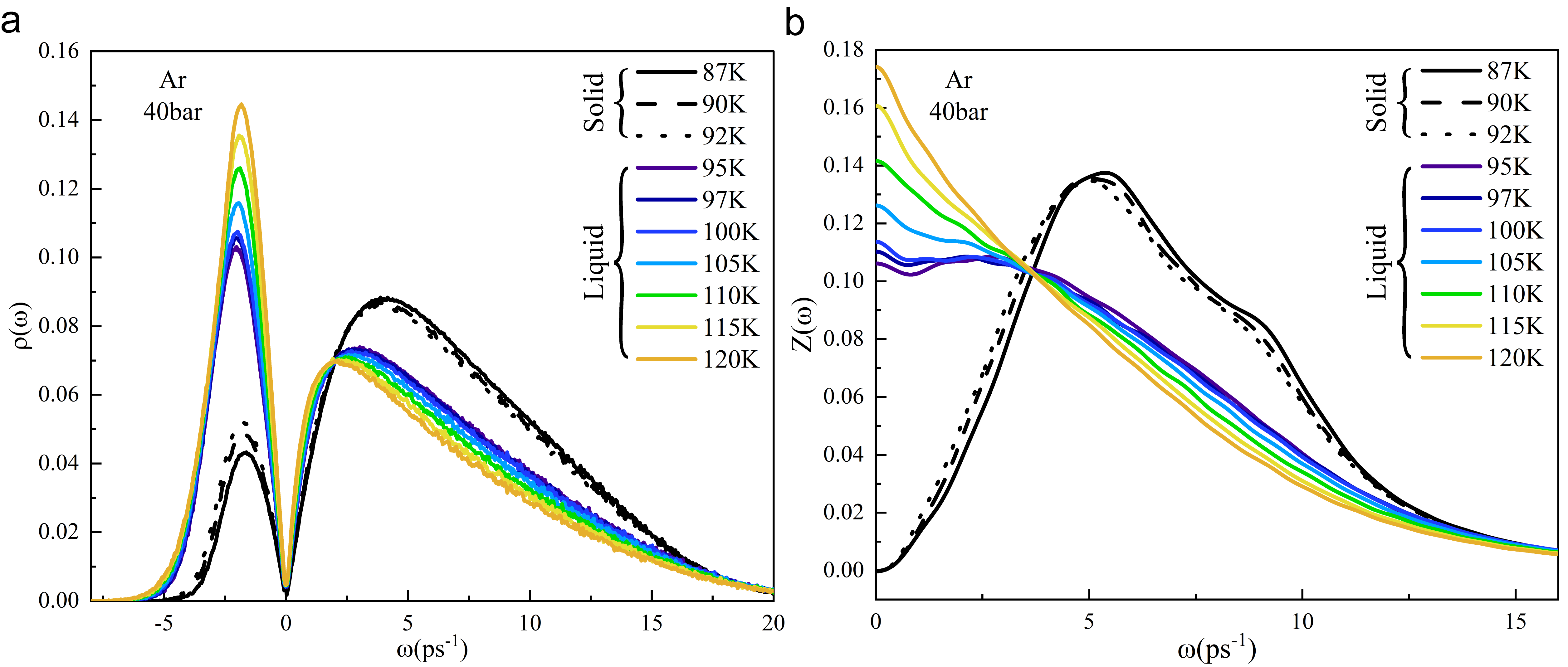}
    \caption{\textbf{(a)} The INM DOS of Argon at $P=40$ bar and different temperatures as defined in Eq.\eqref{inmsdef}. The stable branch is plotted on the positive frequency axes, while the unstable branch on the negative axes using the standard notation described in the main text. The first-order melting transition happens between $92$ K and $95$ K. \textbf{(b)} The density of state function $S(\omega)$ defined in Eq.\eqref{vacfdef} from the Fourier transform of the VACF for the same system. Black lines correspond to the solid phase.}
    \label{fig:3}
\end{figure}

After this initial exploration, we move to study the instantaneous normal modes in the same temperature regime with the idea of comparing the INM density of states with the Fourier transform of the VACF (density of state function), denoted as $S(\omega)$. In this analysis, we have extended the range of temperatures below the melting temperature to explore the solid phase near melting as well. At the same time, we have stopped the analysis at $120$ K without going deep into the gas phase where the differences between the two quantities are too large. All the results in the solid phase are kept in black color to avoid confusion. The color scheme for the other temperatures is kept the same as in Fig.\ref{fig:2}. In panel (a) of Fig.\ref{fig:3}, we show the density of INM using the standard convention of presenting unstable purely imaginary frequencies on the negative frequency axes. Starting from the highest temperature analyzed, $T=120$ K, we can already identify several interesting features. For all temperatures in the liquid phase, both the stable and unstable branches of the INM density of states are linear in frequency at low frequency and they develop a maximum at intermediate frequencies. For the unstable branch, the maximum is located at approximately $2.5$ ps and its location is almost independent of temperature in the whole liquid phase despite its intensity decreases by decreasing temperature. For the stable branch, the location of the maximum moves slightly to higher energies by decreasing the temperature. Below the melting temperature, we observe still a non-vanishing number of unstable modes but a drastic jump in their density. Moreover, the peak in the stable branch jumps quickly to higher energies and approaches an approximately constant value of $5$ ps.

In panel b of Fig.\ref{fig:3} we show the results for the density of state function $S(\omega)$ defined as the Fourier transform of the velocity auto-correlation function $Z(t)$, Eq.\eqref{vacfdef}. In the solid phase (black lines below $95$ K), the density of state function vanishes at zero frequency, as expected for a solid with zero self-diffusion constant. The low-frequency behavior of $S(\omega)$ in the solid phase is compatible with Debye's law and shows a characteristic quadratic scaling $S(\omega) \propto \omega^2$. Moreover, the prefactor in this scaling law grows by increasing the temperature. This can also be rationalized using the Debye model since that prefactor is inversely proportional to the cube of the average speed of sound $\bar v^3$, and the latter clearly decreases by increasing temperature because of the softening of the material. Moving away from the low frequency regime, $S(\omega)$ displays a peak in the solid phase that is located approximately at $\omega=5$ ps$^{-1}$ and it is rather insensitive to temperature. A second shoulder is visible around $\omega=9$ ps$^{-1}$, followed by a monotonic decay.

At approximately $95$ K, the system enters the liquid phase and the behavior of the density of state function $S(\omega)$ changes drastically. First, the zero frequency value of $S(\omega)$ is not anymore zero as a sign of a finite self-diffusion constant that grows with temperature. Moreover, $S(\omega)$ is not anymore monotonic. It first decays monotonically up to a minimum located around $\omega=1$ ps$^{-1}$, then, it grows up to a peak at approximately $2.5$ ps$^{-1}$ and it then decays monotonically again. This behavior is typical of dense liquids, see for example Fig.1 in Ref.~\cite{10.1063/1.1624057}. By increasing further the temperature, both the minimum and the peak at low frequency disappear and the density of state function becomes a monotonically decaying function as expected for dilute liquids and gases. By increasing temperature, $S(\omega)$ gets closer to the Lorentzian function predicted by kinetic theory, a sign of the complete lack of correlations.

We can already anticipate some clear differences and similarities between the stable INM density of states, $\rho(\omega)$ with $\omega>0$, and the density of state function $S(\omega)$. For brevity, we will indicate the positive ($\omega>0$) branch of $\rho(\omega)$ as $\rho_s(\omega)$ in the rest of the manuscript, where the subscript stands for ``stable''.
Let us start from the solid phase near melting (black lines). Even there, $\rho_s(\omega)$ and $S(\omega)$ are profoundly different below $\approx 4$ ps$^{-1}$. In particular, while $\rho_s(\omega)$ is linear in frequency, $S(\omega)$ is quadratic, as Debye's model predicts for solids. The origin of this difference can be found in the existence of a large number of unstable modes, that also persists in the heated crystal away from the zero temperature harmonic limit. These remaining unstable modes  possibly correspond to local anharmonicities that remain present in crystals at finite temperature. Above a certain critical frequency, $\rho_s(\omega)$ and $S(\omega)$ become more similar, showing the same qualitative trend. We notice that the positions of the peak in $\rho_s(\omega)$ and $S(\omega)$ are compatible in the solid phase. We also emphasize that the fact that the INM analysis does not capture correctly the long-time dynamics is not a surprise and it is a direct consequence of the instantaneous approximation adopted. The differences between $\rho_s(\omega)$ and $S(\omega)$ are expected to vanish by decreasing the temperature or increasing the pressure, as discussed in detail in \cite{doi:10.1021/j100189a029}. Below, we will show this trend concretely for another liquid system, Gallium.

By increasing temperature into the liquid phase, even qualitatively, $\rho_s(\omega)$ and $S(\omega)$ are totally different. In particular, $S(\omega)$ acquires a finite zero frequency value, corresponding to the onset of self-diffusion, while $\rho_s(0)$ remains zero. Moreover, $S(\omega)$ becomes non-monotonic, or even monotonically decaying with frequency, while $\rho_s(\omega)$ remains linear in frequency at small frequency.

\begin{figure}[h]
    \centering
    \includegraphics[width=0.8\linewidth]{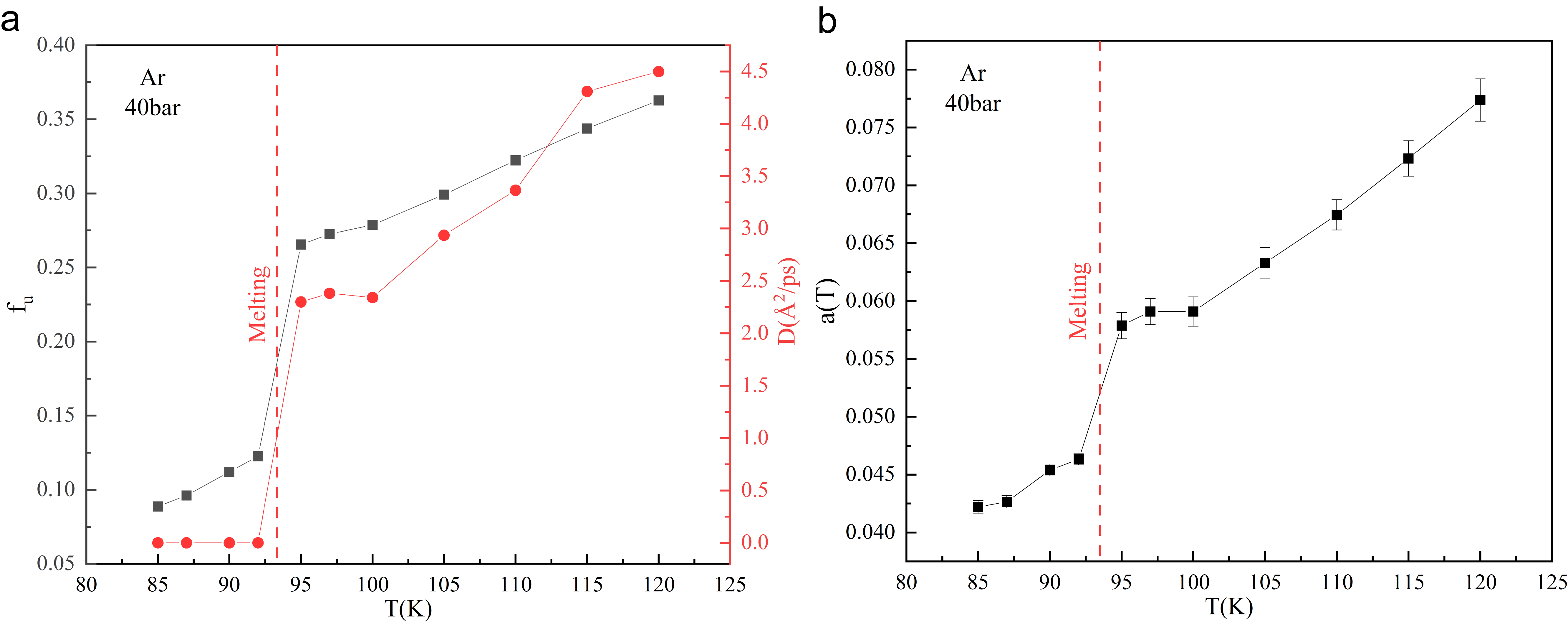}
    \caption{\textbf{(a)} The fraction of unstable modes $f_u(T)$ and the self-diffusion constant $D(T)$ as a function of temperature for Argon at $P=40$ bar. The melting temperature is between $92$ K and $95$ K (dashed red line). \textbf{(b)} The slope of the low-frequency linear scaling of the INM DOS $a(T)$ for the same system and conditions.}
    \label{fig:4}
\end{figure}

We now move to discuss in detail the features of the INM density of states. We are mainly interested in the fraction of unstable modes $f_u(T)$ defined in Eq.\eqref{fudef} and the coefficient of the low-frequency linear scaling $a(T)$, as determined by fitting the INM DOS using Eq.\eqref{atdef}. The fraction of unstable modes is shown as a function of temperature in panel a of Fig.\ref{fig:4}. $f_u$ grows with temperature, as one might expect for a parameter that quantifies the ``fluidity'' of a system. As already anticipated, $f_u$ is not zero in the solid phase near melting but accounts for approximately $10\%$ of the INM. $f_u$ displays a sharp peak at the melting temperature between $92$ K and $95$ K, where it changes abruptly from $\approx 12 \%$ to $\approx 28 \%$. This type of behavior was already observed in the past literature \cite{10.1063/1.474822}. Above melting, in the liquid phase, the fraction of unstable modes grows with temperature following an approximately linear trend. As mentioned above in the section describing the simulation details, it is important to clarify that we did not simulated directly the dynamical melting process. Rather, we have located the melting point by looking at changes in the properties of the system at fixed temperature (see Appendix \ref{ioio} for more details). Our analysis suggests that the system is in a solid state below $T=92$ K but transitions to a liquid state at $95$ K, signifying the occurrence of a phase transition within that range of temperatures.

In panel a of Fig.\ref{fig:4}, we show the self-diffusion constant $D$ as well. Below the melting transition (dashed red line), but not too far from it, the self-diffusion constant becomes very small, as one might expect in crystals \cite{mehrer2007diffusion}. This implies that most of the corresponding non-vanishing unstable modes should not be of diffusive type. In the past, these remaining unstable modes have been associated to so-called ``false barriers'' \cite{10.1063/1.474822} or simply described as localized unstable modes \cite{PhysRevLett.74.936,10.1063/1.3701564}. Above the melting temperature, the diffusion constant becomes large and grows with temperature monotonically. 

In panel b of Fig.\ref{fig:4}, we show the temperature dependence of the linear scaling of the INM DOS, $a(T)$. This scaling coefficient is the same for the stable and unstable parts of the INM spectrum. $a(T)$ grows monotonically with temperature and displays a sharp peak around the melting temperature. Interestingly, its trend between $95$ K and $100$ K seems to follow the behavior of the self-diffusion constant. In particular, in that range of temperatures both $D$ and $a(T)$ are approximately constant before approaching a more steady growth.

\begin{figure}
    \centering
    \includegraphics[width=0.8\linewidth]{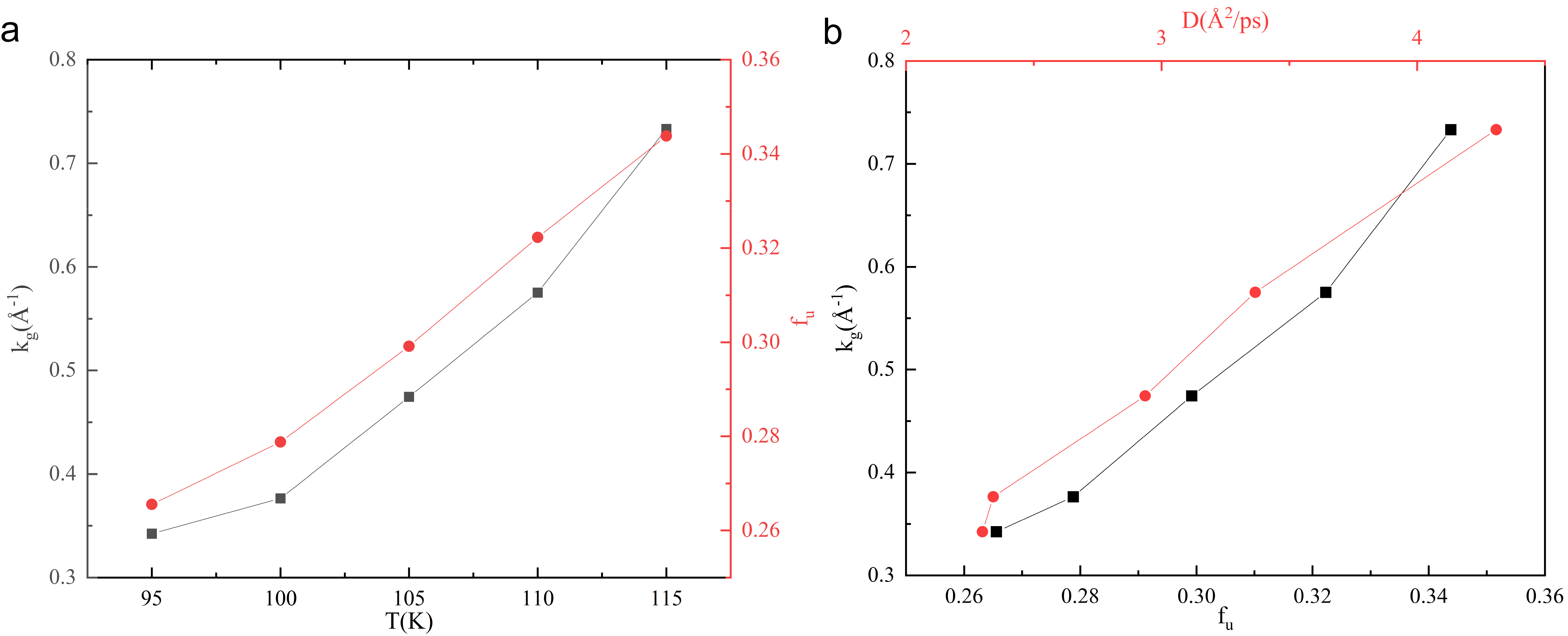}
    \caption{\textbf{(a)} The momentum gap ($k$-gap) of collective shear waves $k_g$ as a function of temperature in the subcritical liquid phase of Argon between $95$ K and $115$ K. The data are taken from Ref. \cite{PhysRevLett.118.215502}. For convenience, the data of the fraction of unstable INM $f_u$ are reproduced in the same panel. \textbf{(b)} The $k$-gap as a function of the fraction of unstable modes $f_u$ and as a function of the self-diffusion constant $D$.}
    \label{fig:5}
\end{figure}

The presence of unstable modes, $f_u \neq 0$, has been often identified as a hallmark of fluidity and a typical feature of liquids. As we have already observed, this is not that simple since a portion of unstable modes survive in the solid phase (at least near to melting) and a more refined classification of the unstable INM is needed to connect them directly with ``fluidity'' (see \textit{e.g.} \cite{PhysRevLett.74.936,10.1063/1.3701564,PhysRevLett.78.2385,PhysRevLett.84.4605,PhysRevE.64.036102,PhysRevE.65.026125,10.1063/1.474822,10.1063/1.474968} for the several criteria proposed). Nevertheless, we find it interesting to compare the fraction of unstable modes and the self-diffusion constant to another measure of ``fluidity'' known as the $k$-gap. As originally advocated by Maxwell and Frenkel \cite{Frenkel1946} (see also fundamental contributions by Zwanzig \cite{10.1063/1.446338}), and later extensively re-discussed \cite{trachenko2015collective,BAGGIOLI20201}, collective shear waves in liquids do not display a linear dispersion relation, as in solids, due to the vanishing of the static shear modulus. On the other hand, the real part of their dispersion (\textit{i.e.}, their energy) follows an approximate square-root form $\mathrm{Re}(\omega) = v \sqrt{k^2-k_g^2}$ where $k_g$ is known in the literature as $k$-gap. This structure has been confirmed in several simulations (\textit{e.g.}, \cite{PhysRevLett.118.215502,PhysRevB.101.214312,PhysRevE.107.055211}) and has been also observed in few experimental systems \cite{PhysRevLett.97.115001,jiang2025experimental}. 

Interestingly, the $k$-gap dispersion relation of shear waves in liquids might also provide an explanation for their linear in frequency DOS at low frequency. Indeed, as explicitly derived in \cite{Trachenko_2023,yu2023unveiling}, a $k$-gap $k_g=1/(2 v \tau_m)$ (with $\tau_m$ the Maxwell relaxation time) implies a DOS of the form
\begin{equation}
    g(\omega)\propto \frac{\omega^2}{v^3} \sqrt{1+\frac{1}{4 \omega^2 \tau_m^2}}.\label{lele}
\end{equation}
At low frequencies, \textit{i.e.} $\omega 
\ll 1/(2 \tau_m)$, the DOS exhibits a linear scaling $g(\omega)\propto \omega/\tau_m$ that is compatible with the findings from the INM analysis. We notice that in that low-frequency regime shear fluctuations are not propagating but rather overdamped (\textit{i.e.}, having imaginary frequency or complex frequency with imaginary part larger than the real). This suggests a connection between linear in frequency DOS, unstable modes and $k$-gap that deserves further attention. We also notice that, despite the INM analysis correctly captures the linear scaling of the DOS at low-frequencies (as predicted as well by $k$-gap theory, Eq.~\ref{lele}), it overestimates the importance of unstable modes. This is specially evident close to the melting temperature where the dispersion of shear waves is almost linear $\omega \approx v k$ (see for example \cite{doi:10.1073/pnas.1006319107,PhysRevB.84.052201,PhysRevLett.102.105502,Hosokawa_2013,Hosokawa_2015}) and therefore the linear contribution is expected to be very small. At this stage, the origin of this problem is not evident but it is likely related to the inherent assumptions of the INM analysis (instantaneous and harmonic).

We reiterate again that the linear scaling of the liquid DOS at low frequency comes from overdamped non-propagating modes. This interpretation is in nice agreement with the above computation from the $k$-gap dispersion, showing that the modes contributing to that linear regime have $\omega  \ll 1/\tau_m$ and, hence, are not propagating wave-like excitations.

As already mentioned, $k_g$ is expected to vanish at the melting transition, below which the collective shear waves reacquire their propagating nature as predicted by hydrodynamics \cite{PhysRevA.6.2401} and elasticity theory \cite{chaikin_lubensky_1995}. This suggests that somehow $k_g$ should relate to the self-diffusion constant and the fraction of ``diffusive'' unstable INM that also vanish in the solid phase. Nevertheless, to the best of our knowledge, how this net of relations is concretely realized has never been investigated.

Taking advantage of the existing data for $k_g$ for subcritical liquid Ar \cite{PhysRevLett.118.215502}, we have attempted a raw comparison among these quantities. The results are shown in Fig.\ref{fig:5}. In panel a, we have presented the data for $k_g$ as a function of temperature in the liquid phase. As expected, $k_g$ grows monotonically with temperature compatible with the idea that $k_g \propto 1/\tau_m$, with $\tau_m$ being the Maxwell time. The latter decreases with temperature and make $k_g$ to increase with $T$. For comparison, in the same panel, we also show the fraction of unstable modes as a function of temperature, to highlight the very similar trend. In panel b of Fig. \ref{fig:5}, we plot $k_g$ as a function of $D$ and $f_u$. In both cases, we observed a very strong linear correlation between these quantities. This is consistent with the idea that all of them are a measure of ``fluidity''. 

In order to find a more robust, and possibly universal relation between these three quantities, few problems need to be solved. First, we notice that despite $f_u$ is a dimensionless quantity, $k_g$ and $D$ are not. Therefore, any universal relation between $f_u$ and $k_g$ for example needs necessarily to involve another dimensionful scale. The same argument holds for a direct relation between $k_g$, which has units of $[1/L]$, and $D$, which has units of $[L^2/t]$. Moreover, as already anticipated, $k_g,D$ are expected to vanish when the solid phase is approached, while $f_u$ does not. In general, it would be interesting to investigate further this point, considering also other more refined measures of $f_u$ that do vanish in the solid phase.

Before concluding, we notice that an approximately linear relation between the diffusion constant $D$ and the $k$-gap $k_g$ could be explained by combining $k$-gap theory with the Stokes-Einstein relation. In fact, the latter implies $D \propto T/\eta= T/(G_\infty \tau_m)$. Using $k_g= 1/(2 v \tau_m)$, one immediately gets $D \propto k_g$, that is compatible with the trend in the simulation data, Fig.~\ref{fig:5}b.

\section{What can we learn from instantaneous normal modes}\label{sec4}
In this Section, we have selected a set of interesting topics and open points surrounding our initial question of \textit{what we can learn from INM}.
\subsection{Fraction of unstable modes}
The fraction of unstable modes defines the percentage of negative eigenvalues (or equivalently imaginary frequencies) in the INM spectrum and it is a function of all thermodynamic variables, such as temperature and pressure.
\begin{figure}[h]
    \centering
    \includegraphics[width=0.8\linewidth]{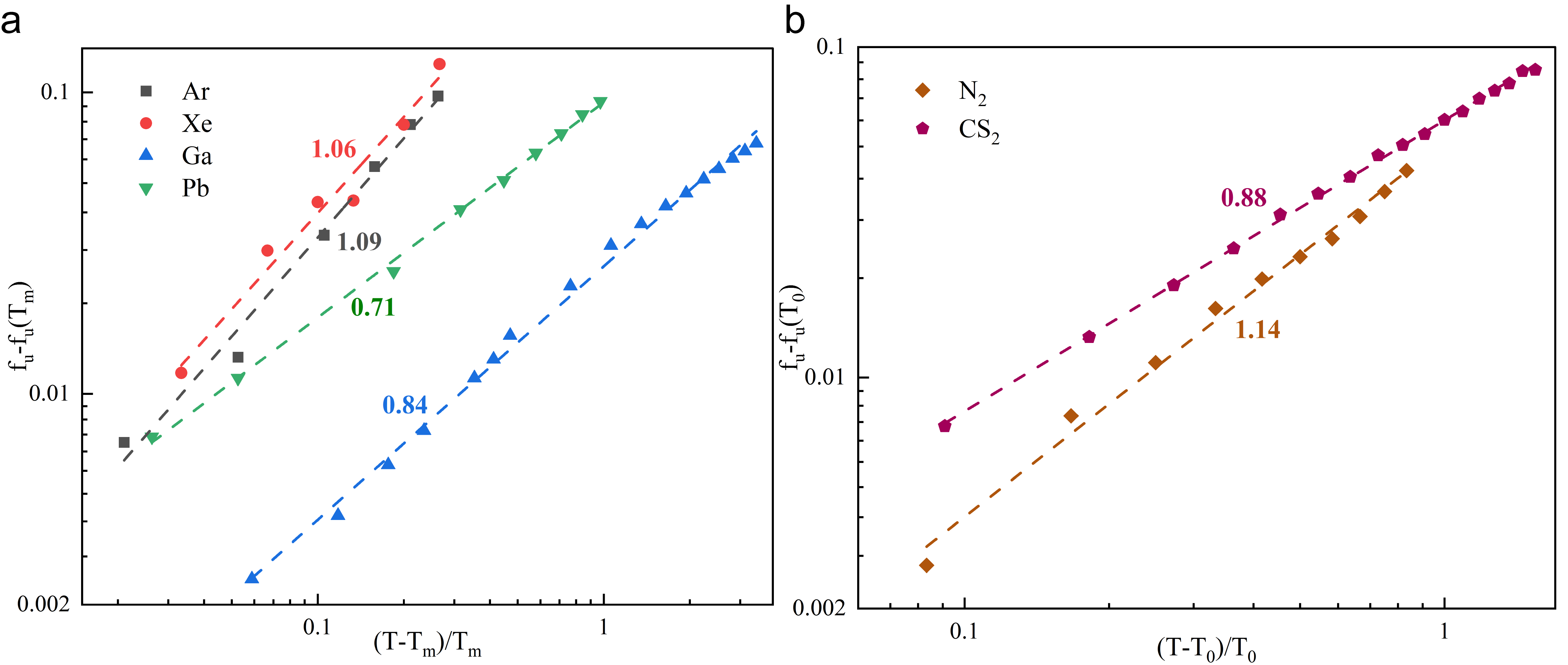}
x    \caption{The fraction of unstable modes as a function of the temperature. The numbers correspond to the power-law $\alpha$ described in the main text. \textbf{(a)} For Ar, Xe, Ga, Pb the data are normalized by their melting temperature $T_m$ and the value of $f_u$ at such temperature. \textbf{(b)} For N$_2$ and CS$_2$ the data are normalized by the lowest temperature available, $T_0$.}
    \label{fig:vvvv}
\end{figure}\\
In all liquids, the fraction of unstable modes increases with temperature. A larger kinetic energy corresponds to a larger probability of exploring regions of the potential landscape with locally negative curvature, such as saddle points or maxima, moving the dynamics away from the bottom of the local minima. Despite this might seem natural from a physical perspective, how $f_u$ increases with temperature is much less trivial. In order to provide valuable information regarding the temperature trend of $f_u$, we have collected all our results in Fig.\ref{fig:vvvv}. In panel a, we show $f_u$ for Ar, Xe, Ga and Pb. For a better comparison, we normalized the data using the melting temperature of each of these liquids $T_m$ and the corresponding value of $f_u$ at melting. We observe that $f_u(T)$ is well described in all cases by a power-law function $f_u(T)-f_u(T_m)\sim \left(T-T_m\right)^\alpha$, where nevertheless $\alpha$ depends on the specific liquid chosen. More precisely we find that $\alpha$ is $1.09,1.06,0.84,0.71$ for Ar, Xe, Ga and Pb respectively. For N$_2$ and CS$_2$ we do not have data up to the melting temperature and we therefore decided to normalize the data by the minimum temperature available, $T_0$, and show the corresponding results in a separate panel, panel b in Fig.\ref{fig:vvvv}. There, we find again a trend consistent with a single power-law, where the power $\alpha$ is $1.14$ for N$_2$ and $0.88$ for CS$_2$.

It is noteworthy that the power-law $\alpha$ is close to $1$ for all the liquids considered. This is in good agreement with the predictions from the random energy model (see \textit{e.g.} \cite{keyes-2005}), according to which $f_u(T)=\frac{T}{\delta 2\pi} e^{-\delta^2/(4 T^2)}$
where $\delta$ is an $\mathcal{O}(1)$ parameter describing the width of the energy distribution. We have tried to fit the data with the expression from the random energy model with a single free parameter $\delta$ and found that it does not capture correctly the trend of the numerical data.

We notice that the exponent $\alpha$ measures the rate at which a given liquid develops unstable modes by increasing the temperature. This exponent is possibly related both to dynamical properties of the liquids but, more fundamentally, to the dynamics of its potential energy landscape. In this direction, it would be interesting to explore which properties of a liquid determines determine such exponent and whether the latter displays any strong correlation with any other measurable quantity.

\subsection{Low frequency linear scaling}
As already anticipated, another universal feature of the INM density of states is its linear low-frequency behavior of the form
\begin{equation}\label{lili}
   \rho(\omega)= a(T) |\omega|+\dots
\end{equation}
This behavior can be derived using several theoretical arguments \cite{doi:10.1063/1.457564,doi:10.1063/1.465563,doi:10.1063/1.467178,doi:10.1073/pnas.2022303118,doi:10.1073/pnas.2119288119,doi:10.1063/1.479810,PhysRevE.55.6917,doi:10.1063/1.463375} and it is directly related to the presence of unstable modes with imaginary frequency. In all the systems considered in this manuscript, that include Ar, Xe, Ga, Pb, N$_2$ and CS$_2$, the INM DOS follow the linear form presented in Eq.\eqref{lili}. In general terms, much less is known about the temperature dependence of the parameter $a(T)$ that characterizes the slope of this linear trend. 

In a series of works \cite{keyes1994unstable,doi:10.1063/1.479810} (see \cite{keyes-2005} for a review), Keyes and collaborators proposed a closed-form formula for this parameter that, under some reasonable approximations, takes the following form
\begin{equation}\label{slopetheory}
    a(T)\propto \left[f_u^{\text{max}}-f_u(T)\right]\,e^{-\langle E\rangle/k_B T}\,.
\end{equation}
Here, $f_u^{\text{max}}$ is the maximum value for the fraction of unstable modes, and depends on the details of the potential landscape. On the other hand, $\langle E \rangle$ is an average energy scale that is supposedly related to hopping processes and the anharmonic nature of the potential landscape. As seen in the previous Section, the fraction of unstable modes grows with temperature. This implies that the first term in Eq.\eqref{slopetheory} is a term that decreases with temperature. The other contribution to Eq.\eqref{slopetheory} is an exponential growth of the scaling upon increasing the temperature, whose growth rate is controlled by the energy scale $\langle E \rangle$. Because of the competition between these two terms, in general Eq.\eqref{slopetheory} predicts that $a(T)$ grows/decreases with temperature depending on which of the terms therein dominates. These two behaviors have been both observed in simulations. For a Lennard Jones liquid \cite{keyes1994unstable}, the energy scale $\langle E \rangle$ is very low; hence, the first term in Eq.\eqref{slopetheory} dominates the behavior of $a(T)$ that decreases with temperature. The same is observed for bulk polymers \cite{doi:10.1063/1.5127821}. On the contrary, for many other systems, including CS$_2$ \cite{doi:10.1063/1.479810} and water \cite{jin2024temperature}, $a(T)$ grows with temperature following a form compatible with the exponential term in Eq.\eqref{slopetheory}.

We notice that a slope of the form $a(T) \propto \exp\left(\langle E \rangle/k_B T\right)$ could be also explained from gapped shear waves ($k$-gap theory \cite{BAGGIOLI20201}). Indeed, as pointed out in \cite{Trachenko_2023,yu2023unveiling}, gapped shear waves would contribute a term of the form $g(\omega)\propto \omega/\tau_m$. A relaxation time $\tau_m \propto \exp \left(\langle E \rangle/k_B T\right)$, that decreases exponentially with temperature, would immediately imply such a temperature dependence for $a(T)$. Nevertheless, this argument would not explain the scenarios (\textit{e.g.}, LJ liquids \cite{keyes1994unstable}) where $a(T)$ decreases with temperature.

\begin{figure}[h]
    \centering
    \includegraphics[width=1\linewidth]{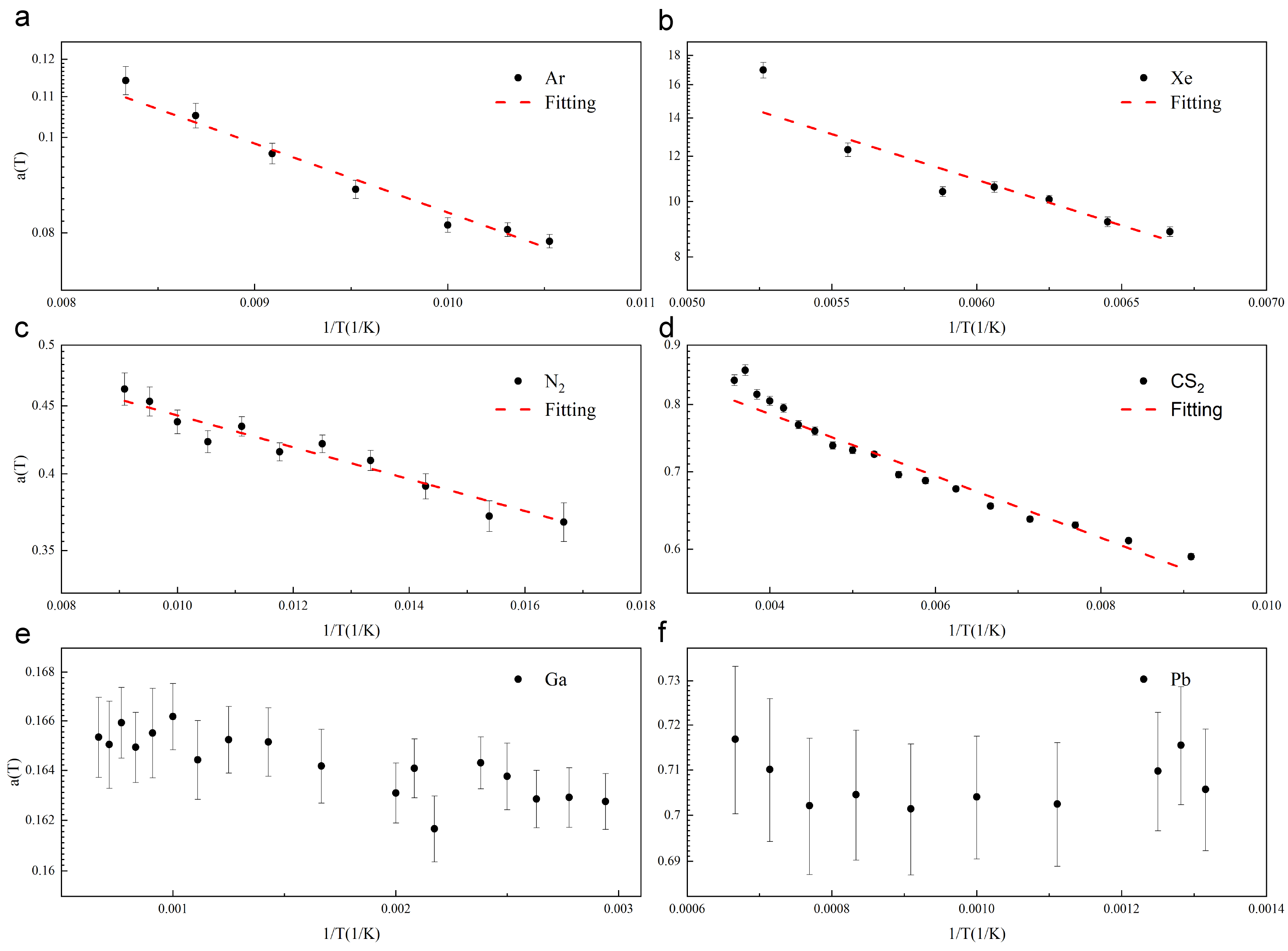}
    \caption{The INM slope extracted from the experimental data using Eq.\eqref{lili} as a function of the inverse temperature $1/T$ for Ar \textbf{(a)}, Xe \textbf{(b)}, N$_2$ \textbf{(c)}, CS$_2$ \textbf{(d)}, Ga \textbf{(e)} and Pb \textbf{(f)}. The red dashed line in panels (a)-(d) is the result of an exponential fit $a(t)\propto \exp (-\langle E \rangle / k_B T)$. For all systems, the data points are above the melting temperature.}
    \label{fig:vvvvw}
\end{figure}

We have analyzed the slope parameter $a(T)$ for our six different liquid systems. For Ar, Xe, N$_2$ and CS$_2$ we observe that the slope $a(T)$ clearly increases with temperature. Moreover, for these systems the temperature dependence of the slope is compatible with an Arrhenius-like exponential form $a(t)\propto \exp (-\langle E \rangle / k_B T)$, where the activation energy is given by:
\begin{align}
    &\text{Ar:}\,\langle E \rangle = 13.88 \,\text{meV} \,,\qquad  \text{Xe:}\, \langle E \rangle= 31.66\, \text{meV}\,,\qquad  \text{N$_2$:}\,\langle E \rangle= 2.384 \,\text{meV}\,, \qquad 
       &\text{CS$_2$:}\,\langle E \rangle= 5.301 \,\text{meV}\,.
\end{align}
We notice that the reported value of $\langle E \rangle \approx 13 \,\text{meV}$ for water \cite{jin2024temperature} is of the same order of magnitude of the values found above. 
As we will see in the next Section, these activation energies are consistently smaller than those governing the Arrhenius behavior of the corresponding diffusion constants. For example, for $N_2$ the activation energy characterizing the slope $a(T)$ is about $20$ times smaller than the one describing $D(T)$. This implies that, despite the same functional form, it is very unlikely that the microscopic origin of it is the same for the slope $a(T)$ and the diffusion constant $D(T)$. Nevertheless, we notice that all the normal modes analysis, including therefore the determination of $a(T)$, is performed using the harmonic approximation. On the contrary, the determination of $D(T)$ does not rely on this approximation. It is therefore conceivable (similar arguments have been made in \cite{jin2024temperature} by comparing the INM slope and the experimental slope) that the strong liquid anharmonicities are the responsible for this difference.

Finally, we clarify that, to the best of our knowledge, no existing theoretical framework to determine from first principles $\langle E \rangle$ exists. Our numerical results could be useful in this direction to achieve such a theoretical model and directly test it with simulations.

On the other hand, as shown in panels (e)-(f) in Fig.\ref{fig:vvvvw}, the temperature dependence of the slope in Ga and Pb shows a completely different trend. In particular, within the accuracy of our numerics, the slope seems rather independent of temperature and shows an approximate constant behavior. At the moment, we do not have any explanation for this behavior but we can already offer two insights. (I) The numerical data are certainly incompatible with an exponential behavior $a(t)\propto \exp (-\langle E \rangle / k_B T)$ as found for the other simpler liquids. (II) Both Ga and Pb have in common the fact that they are liquid metals, providing a possible hint to rationalize this different trend.

\subsection{Self-diffusion constant}
One of the initial motivations to introduce and study instantaneous normal modes was the idea of reconstructing the liquid self-diffusion constant in terms of liquid normal modes \cite{10.1063/1.446338}. Keyes and collaborators have proposed that the self-diffusion constant in liquids could be related to the unstable instantaneous normal modes \cite{10.1063/1.458192}, and that a somehow simple relation between $D$ and $f_u$ could exist,
\begin{equation}\label{dslo}
    D(T)= \text{const.}\, \langle \omega_u\rangle f_u.
\end{equation}
Here, $\langle \omega_u\rangle$ is the average unstable frequency. We notice that in Eq.\eqref{dslo}, ``const.'' is not a dimensionless number but it is assumed to be temperature independent.

Despite INM describes diffusion in liquids well \cite{keyes1997instantaneous,keyes1994unstable,10.1063/1.458192,10.1063/1.469947}, several open questions remain. First, it is by now well-known that the total fraction of unstable modes cannot be a direct source for self-diffusion since $D=0$ in solids but $f_u$ is not. As already mentioned above, this has motivated a large number of studies to find out which subset of $f_u$ is actually responsible for self-diffusion (\textit{e.g.}, \cite{PhysRevE.64.036102,PhysRevLett.78.2385} and discussions therein). Moreover, in several instances (\textit{e.g.}, CS$_2$) it has been shown that Eq.\eqref{dslo} does not hold and a ``stretched'' version of Eq.\eqref{dslo} has to be considered (see \textit{e.g.} \cite{doi:10.1063/1.479810}),
\begin{equation}\label{dslo2}
    D(T) \sim \left( \langle \omega_u\rangle f_u\right)^\gamma.
\end{equation}
We have computed the temperature dependent self-diffusion constant $D(T)$ for our six liquid systems. The behavior of $D$ as a function of temperature is presented in Arrhenius form in \ref{app:D}. All data are consistent with an Arrhenius behavior,
\begin{equation}
    D(T) \propto e^{-\mathcal{E}/T}.
\end{equation}
Our numerical fits give:
\begin{align}
    &\text{Ar:}\qquad \mathcal{E}=34.01  \text{meV}\,,\qquad \qquad \text{Xe:}\qquad \mathcal{E}=60.85 \text{meV}\,,\qquad \qquad \text{N$_2$:}\qquad \mathcal{E}=45.58 \text{meV}\,, \nonumber\\
       &\text{CS$_2$:}\qquad \mathcal{E}=41.17 \text{meV}\,,\qquad \qquad \text{Ga:}\qquad \mathcal{E}=88.82 \text{meV}\,,\qquad \qquad \text{Pb:}\qquad \mathcal{E}=170.4 \text{meV}\,. \label{diffidiffi}
\end{align}

\begin{figure}[h]
    \centering
    \includegraphics[width=0.8\linewidth]{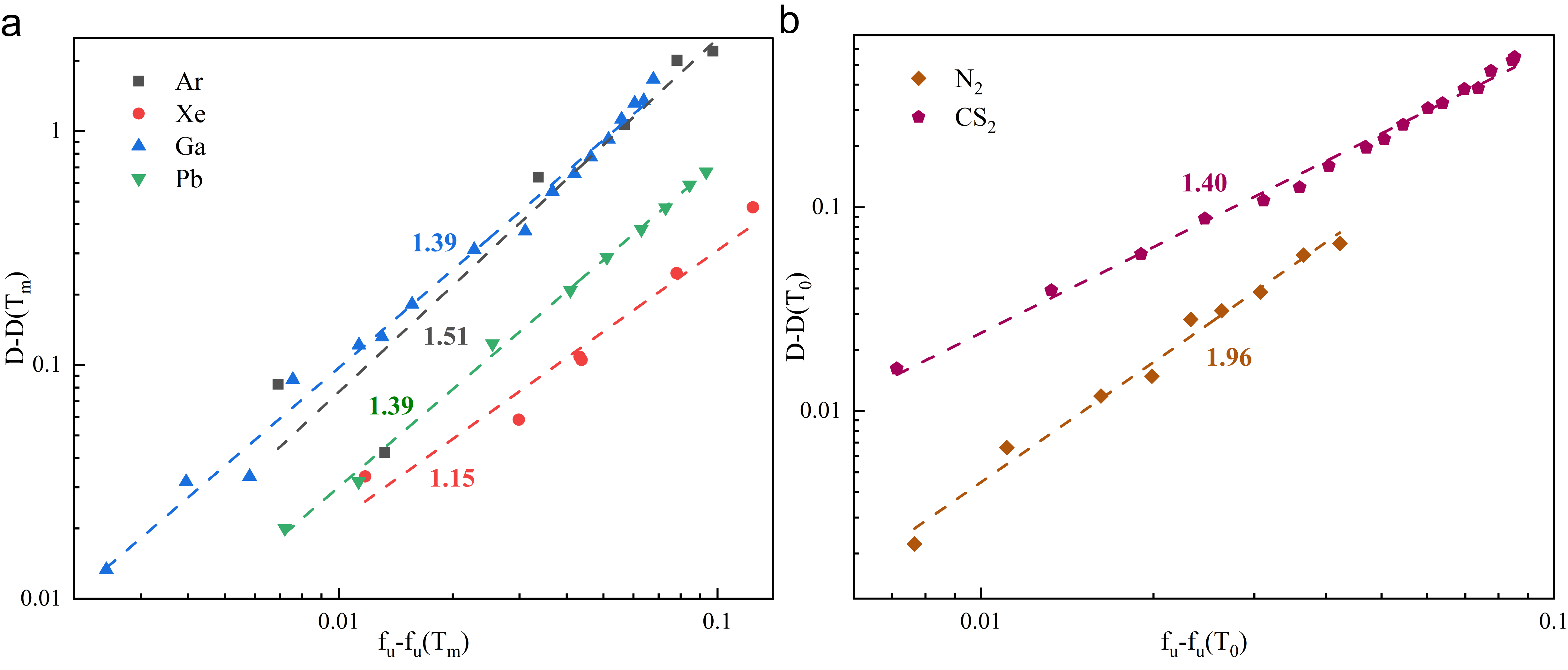}
    \caption{The self-diffusion constant as a function of the fraction of unstable modes $f_u$. The numbers correspond to the power-law $\beta$ described in the main text. \textbf{(a)} For Ar, Xe, Ga, Pb the data are normalized by their melting temperature $T_m$ and the value of $D$ at such temperature. \textbf{(b)} For N$_2$ and CS$_2$ the data are normalized by the lowest temperature available, $T_0$.}
    \label{fig:vvvvw3}
\end{figure}

As already emphasized in the previous Section, the activation energy $\mathcal{E}$ governing the Arrhenius behavior in $D(T)$ is consistently larger than the one describing the increase of the INM slope in the corresponding systems. Not only that, but even the trend of the values between the different liquids is not the same. For example, $\langle E \rangle_{\text{Ar}} >\langle E \rangle_{\text{CS}_2}$ but $\mathcal{E}_{\text{Ar}} <\mathcal{E}_{\text{CS}_2}$. This implies that either the two energy scales reflect very different properties of a liquid or anharmonicities, that are neglected in the determination of $\langle E \rangle$, play a fundamental role here and are the origin of this discrepancy. 

In order to investigate further the relation between the self-diffusion constant $D$ and the fraction of unstable modes, in Fig.\ref{fig:vvvvw3} we have plotted $D$ as a function of $f_u$. In order to compare the different systems, we have normalized both quantities by subtracting their value at the melting temperature $T_m$ (for Ar, Xe, Ga and Pb) or at the minimum temperature available $T_0$ (for N$_2$ and CS$_2$). For all the systems considered, we observe a power law scaling, with the power $\beta$ ranging from $1.15$ in Xe to $1.96$ in CS$_2$. We notice that most of the powers obtained by fitting are not far from the value of $1$, that would be consistent with Eq.\eqref{dslo} and the analysis in \cite{doi:10.1063/1.479810}. Nevertheless, we also observe important deviations and we do not find any universal scaling.

\subsection{Low temperature solid-like limit}
Let us consider a liquid. Its INM DOS is linear in the frequency. On the other hand, Debye taught us that for an ideal crystal, the corresponding vibrational density of states has a quadratic behavior, known as Debye's law \cite{kittel2018solid}. The contrast between these two opposite trends has been observed experimentally in selenium \cite{PhysRevLett.63.2381}, heavy water \cite{DAWIDOWSKI2000247} and water \cite{jin2024temperature}. Less is known about how the INM DOS interpolates between these two behaviors by lowering the temperature and approaching the ideal crystalline phase (see nevertheless \cite{10.1063/1.474968} where this transition is observed when only pure translation INM are considered). As already mentioned, the presence of unstable modes (no matter how few) implies that a linear scaling appears in the INM DOS. Therefore, as soon as unstable mode are present in the spectrum and $f_u\neq 0$, the INM DOS will be linear at low enough frequencies. By decreasing temperature, the fraction of unstable modes diminishes and the linear scaling is expected to get confined to very low frequency and eventually give room to the Debye scaling, that is allowed to appear only when $f_u=0$.
\begin{figure}[h]
    \centering
    \includegraphics[width=\linewidth]{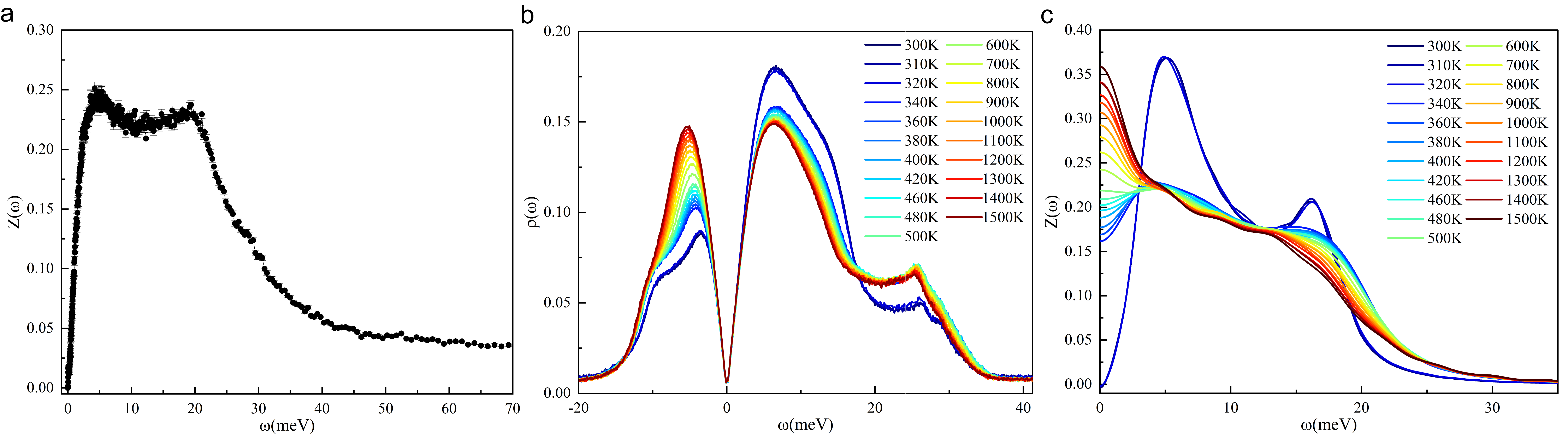}
    \caption{\textbf{(a)} The experimental density of states for liquid gallium at $500$ K, We thank Dehong Yu and the Authors of Ref. \cite{doi:10.1021/acs.jpclett.2c00297} to provide us the experimental data. \textbf{(b)} The INM DOS of Gallium simulated at different temperatures. The first-order melting temperature is between 320 K and 340 K. \textbf{(c)} The density of state function $S(\omega)$ derived from the Fourier transform of the VACF for the same simulated Gallium system.}
    \label{fig:7}
\end{figure}

In order to investigate this point, we have chosen Gallium as a benchmark example. As a reference, in Fig.\ref{fig:7}, we present the experimental data for the density of states at $500$ K from Ref.\cite{doi:10.1021/acs.jpclett.2c00297} together with our simulation results for the INM density of states $\rho(\omega)$ and the density of states function $S(\omega)$. Among the various features emerging from this comparison, we notice that at low temperature (\textit{e.g.}, $T=300$ K), $S(\omega)$ shows already a clear Debye scaling at low frequency and a vanishing diffusion constant ($S(0)=0$). On the contrary, the stable branch of the INM DOS is still strongly linear at low frequency, consistent with a large number of unstable mode still present at such a temperature. This difference is caused by the presence of non-diffusive unstable modes that survive in the solid phase. These modes are possibly a spurious effect of the INM computation and can be removed using various methods. We defer a detailed study of the nature of these modes for future work.
\begin{figure}[h]
    \centering
    \includegraphics[width=0.7\linewidth]{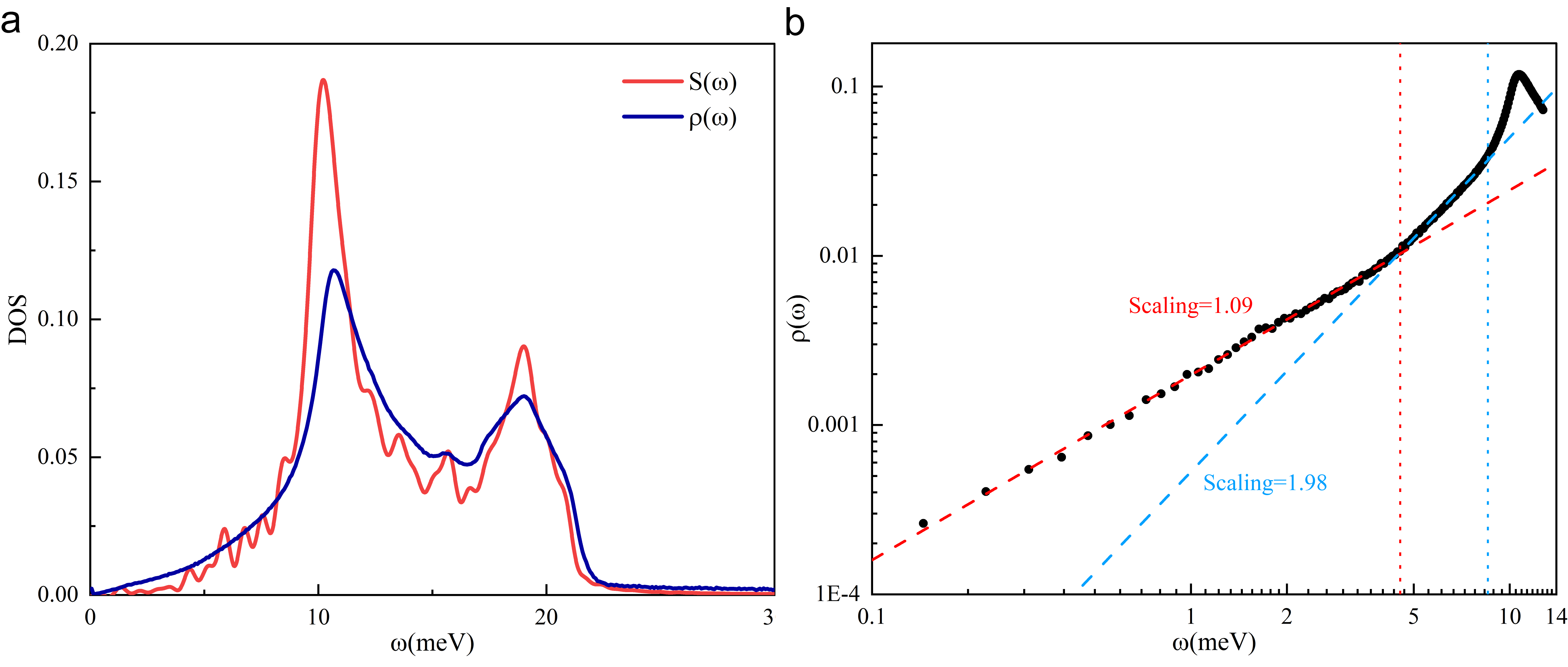}
    \caption{Simulated Gallium at $1$ K. \textbf{(a)} The comparison between the stable branch of the INM DOS $\rho(\omega)$ and the density of states function $S(\omega)$ derived from VACF. \textbf{(b) } Zoom of the low frequency region of the stable branch of the INM DOS. Red and blue dashed lines guide the eyes towards the Debye scaling $\omega^2$ (blue) and the linear scaling (red). The corresponding vertical dotted lines mark the regions where these scalings appear.}
    \label{fig:8}
\end{figure}

In order to test the emergence of the Debye scaling in the INM spectrum, we have lowered the temperature to a value of $T=1$ K. The results of this numerical analysis for $\rho(\omega)$ and $S(\omega)$ are presented in Fig.\ref{fig:8}.  The density of state function $S(\omega)$ presents a Debye scaling $S(\omega) \propto \omega^2$ up to approximately $8$ meV and it exhibits two distinct peaks at $\approx 10$ meV and $18$ meV. Moving to the INM analysis, at $T=1$ K, we observe that the number of unstable modes in the INM spectrum becomes very small. The shape of $\rho(\omega)$ is now very consistent with the density of state function $S(\omega)$, as shown in panel (a) of Fig.\ref{fig:8}. The INM analysis is able to capture well the position of the two aforementioned peaks, despite their width is slightly larger and intensity smaller. This seemingly good agreement between the two quantities is a consequence of the fact that the harmonic approximation becomes more and more reliable at low temperature. In fact, at exactly $T=0$, the two computations should give exactly the same results, up to numerical precision.

In order to check this point further, in panel (b) of Fig.\ref{fig:8}, we have zoomed the stable part of the INM spectrum of Gallium at $T=1$ K and presented it in a log-log scale to highlight possible power-law scalings. The data present a clear linear scaling up to approximately $4.5$ meV, that is a consequence of the remaining unstable modes (and anharmonicities). On the other hand, around $\approx 4.5$ meV we observe a clear crossover to a larger power that is consistent with the Debye scaling $\rho(\omega) \propto \omega^2$ and that persists to $\approx 8$ meV. This reveals the approach of the solid-like behavior to low frequency by decreasing the temperature. By decreasing further $T$, one expects that the crossover frequency will move towards the origin, and finally vanish at $T=0$ when $f_u=0$. We notice that an alternative method to observed this trend, and the emergence of the Debye scaling, consists in nano-confining the liquid, as experimentally performed in \cite{yu2023unveiling}. We also notice that the presence of unstable modes in finite temperature (amorphous) solids has been verified experimentally, see for example \cite{vaibhav2024experimental}.

\subsection{Zero eigenvalue cusp singularity}
Following previous observations \cite{sastry2001spectral,taraskin2002disorder}, Refs. \cite{doi:10.1073/pnas.2119288119,Mossa2023} have recently pointed out that the frequency representation of the INM density of states $\rho(\omega)$ might hide important information that are contained in the corresponding eigenvalue distribution $\upsilon(\lambda)$. Their arguments are supported by numerical simulations in a soft-sphere liquid model (data reproduced in panel (a) of Fig.\ref{fig11}). Since $\lambda =\omega^2$, the eigenvalue distribution and the INM DOS are simply related via the Jacobian,
\begin{equation}
    \rho(\omega) = 2 |\omega| \upsilon(\lambda = \omega^2).
\end{equation}
Mathematically, it appears therefore unreasonable that one of the two quantities contain more information than the other. Nevertheless, in this Section we entertain this possibility and analyze in detail the arguments of \cite{doi:10.1073/pnas.2119288119,Mossa2023} in a large set of liquid models. First, we would like to point out that the statement that the INM density of states is universally linear at low frequencies is equivalent to the statement that $\rho(0)\neq 0$, that is certainly true. In other words, the presence of a cusp at $\lambda=0$ is not incompatible with a low-frequency form $\rho(\omega)=a(T) \omega +\dots$, observed in many instances.

\begin{figure}[h]
    \centering
    \includegraphics[width=\linewidth]{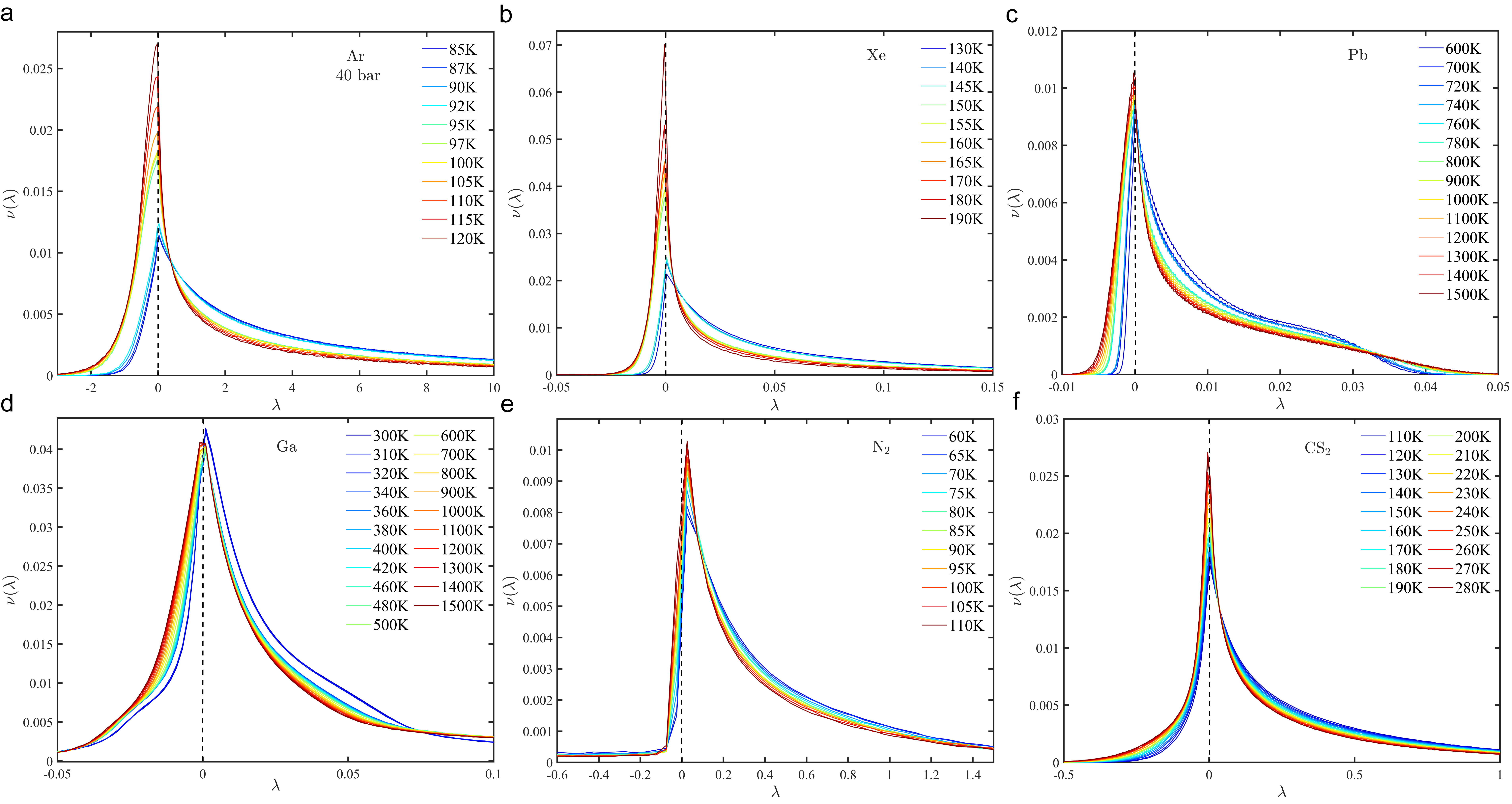}
    \caption{The eigenvalue distribution $\upsilon(\lambda)$ for different temperatures. From top left to bottom right: Ar (40 bar) \textbf{(a)}, Xe ($1.01$ bar) \textbf{(b)}, Pb($1.01$ bar) \textbf{(c)}, Ga($1.01$ bar) \textbf{(d)}, N$_2$ ($70.92$ bar) \textbf{(e)} and CS$_2$ ($1.01$ bar) \textbf{(f)}.}
    \label{fig:88b}
\end{figure}

Here, we have obtained the eigenvalue distribution $\upsilon(\lambda)$ for six different liquids (Ar, Xe, Pb, Ga, N$_2$ and CS$_2$). The results are shown in a large range of temperatures in Fig.\ref{fig:88b}. The cusp-like singularity at $\lambda=0$ is present in all the systems considered. The temperature dependence of the value of $\upsilon(\lambda)$ at the cusp, $\lambda=0$ does not bring any new information since $\upsilon(0)=a(T)/2$, where $a(T)$ is the slope of the low-frequency regime in the frequency representation $\rho(\omega)$. The asymmetry character of the cusp-like singularity is more interesting and relates to the different behavior of the stable and unstable INMs upon moving temperature.

In order to characterize these features in a more quantitative way, we resort to the empirical function proposed in \cite{Mossa2023},
\begin{equation}\label{empi}
    \upsilon(\lambda)=\frac{e^{-a \lambda } (c_0+c_1 \lambda )+e^{a \lambda }
   (c_2+c_3 \lambda )}{e^{-a \lambda }+e^{a \lambda }}
\end{equation}
where $c_1,c_3$ are referred to as ``slopes'' of the cusp-singularity. More precisely, $c_1$ is the slope of the stable branch and $c_3$ that of the unstable one.

\begin{figure}[h]
    \centering
    \includegraphics[width=0.75\linewidth]{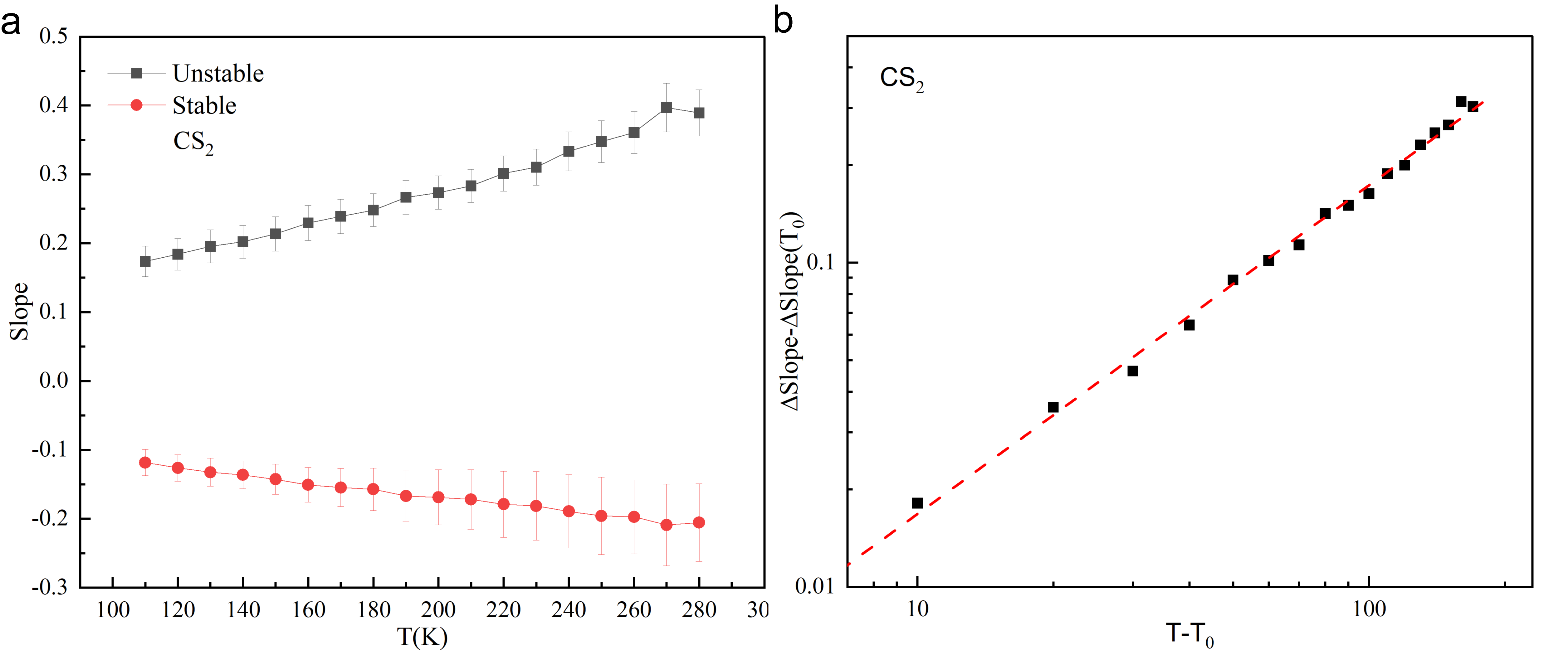}
    \caption{\textbf{(a)} The temperature dependence of the ``slopes'' of the eigenvalue distribution $\upsilon(\lambda)$ for CS$_2$ extracted using the empirical formula Eq.\eqref{empi}. \textbf{(b)} The slope difference as a function of temperature. The dashes line guides the eyes towards the linear scaling observed.}
    \label{fig:8t}
\end{figure}

In panel (a) of Fig.\ref{fig:8t}, we show the temperature behavior of the fitted stable (red) and unstable (black) slopes as a function of temperature for the case of CS$_2$. The slopes are extracted from the numerical data using Eq.\eqref{empi}. Similar results are obtained for the different liquids shown in Fig.\ref{fig:88b}. We observe that the positive slope of the unstable branch grows with temperature while the negative one, corresponding to the unstable part of the spectrum, moves towards larger negative values by increasing $T$. This trend is opposite of what reported in \cite{doi:10.1073/pnas.2119288119,Mossa2023} for the soft sphere model. The reason for this difference is ultimately the opposite behavior of $\upsilon(0)$ with temperature in the soft sphere model. In particular, as evident from Fig.1 in \cite{doi:10.1073/pnas.2119288119}, $\upsilon(0)$ in the soft sphere model decreases by increasing temperature. In our language, this corresponds to an INM slope $a(T)$ decreasing with temperature, as for LJ liquids, but opposite to our more realistic liquid systems. In fact, one could argue that the soft-sphere model is not a realistic model for liquids since it does not present a cohesive state. The relevance of the INM results in the soft-sphere model is therefore arguable, specially when compared to the more realistic cases considered in this work, that present opposite behavior.

This contrast is then reflected in the temperature dependence of the difference of the slopes, that is presented in panel (b) of Fig.\ref{fig:8t}. There, we observe that $\Delta_{\text{slope}}$ increases with temperature, following a trend approximately consistent with a linear power-law scaling. This is again very different from the $T^{-2/3}$ behavior observed in the soft sphere model \cite{doi:10.1073/pnas.2119288119,Mossa2023}, not only for the precise power involved in the scaling, but even in the qualitative trend as well. Our findings imply that the $T^{-2/3}$ law found in \cite{doi:10.1073/pnas.2119288119,Mossa2023} is not universal and its value and significance should be revisited.

Finally, we would like emphasize that the theoretical model presented in \cite{doi:10.1073/pnas.2119288119,Mossa2023} is not the only one able to predict the presence of a cusp-like singularity in the eigenvalue distribution of liquids. Indeed, a much older model known as the ``soft potential model of liquids'' (SPM) \cite{PhysRevE.55.6917,10.1063/5.0158089} already captures such a feature. We provide a direct proof of this statement in panel b of Fig.\ref{fig11}. The presence of a cusp singularity at $\lambda=0$ is evident. For completeness, in the inset we also show the INM DOS that shows the aforementioned linear in frequency regime despite the cusp.
\begin{figure}[ht!]
    \centering
    \includegraphics[width=0.47\linewidth]{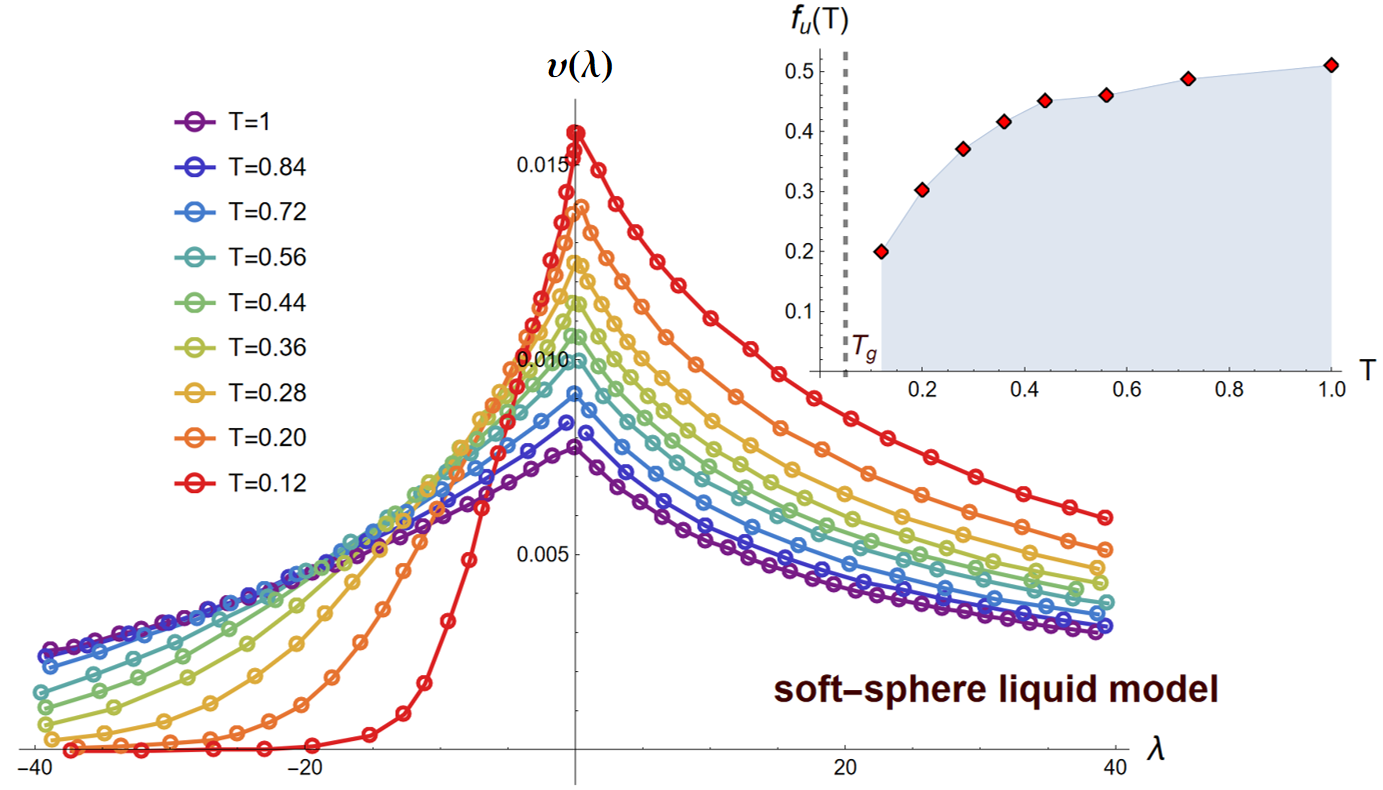}\quad
   \includegraphics[width=0.47\linewidth]{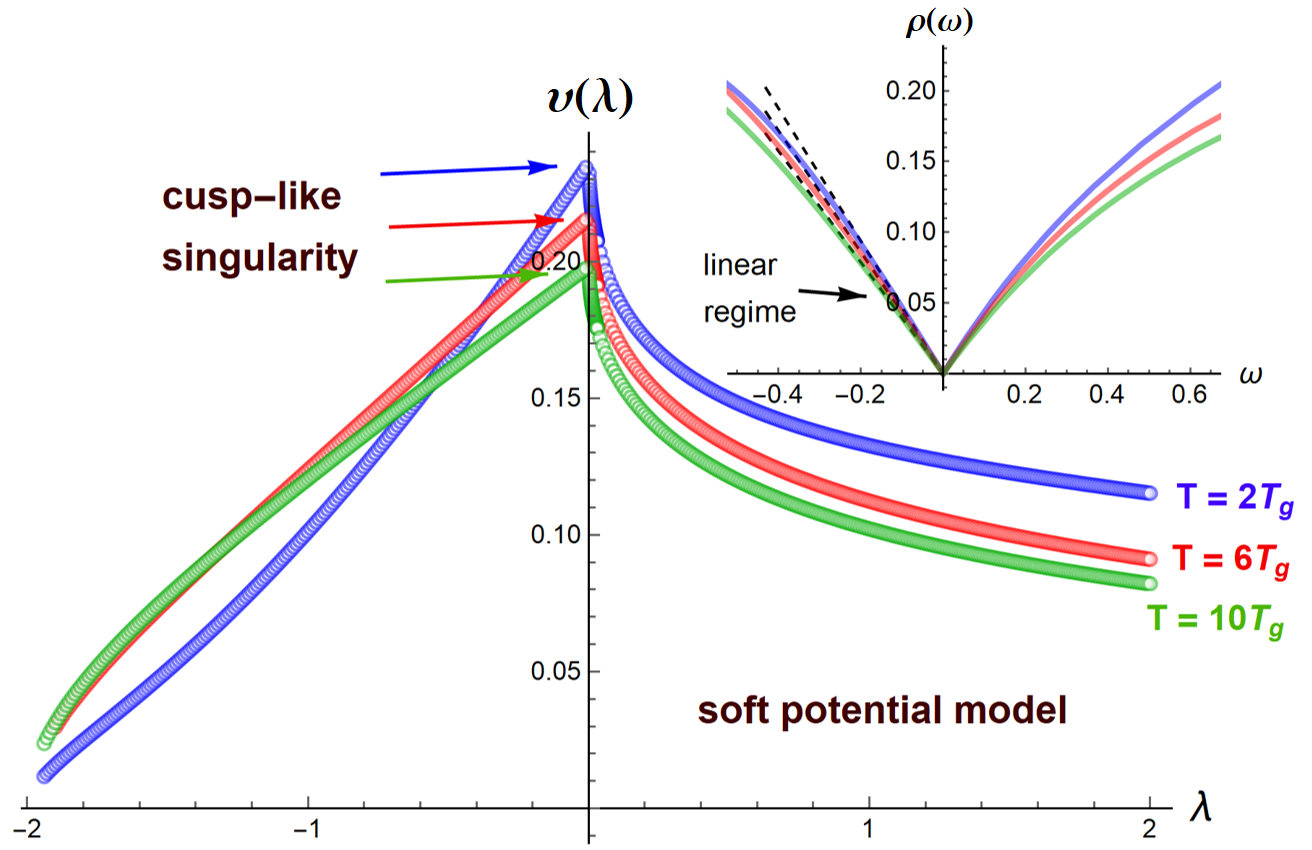}
    \caption{\textbf{(a)} The eigenvalue distribution $\upsilon(\lambda)$ for the soft-sphere liquid model presented in \cite{Mossa2023} and the corresponding fraction of unstable mode $f_u$ as a function of temperature in reduced LJ units. $T_g$ is the glass temperature of the corresponding system. \textbf{(b)} The eigenvalue distribution $\upsilon(\lambda)$ obtained from the theoretical soft potential model of liquids \cite{PhysRevE.55.6917,10.1063/5.0158089} by converting the calculated $\rho(\omega)$ to $\upsilon(\lambda)$ with the Jacobian relation. The inset shows the original INM DOS $\rho(\omega)$ with the dashed black lines indicating the universal linear regime $\rho(\omega) \propto \omega$ which emerges clearly despite the presence of the cusp-like singularity in $\upsilon(\lambda)$.}
    \label{fig11}
\end{figure}\\

In this direction, it would be interesting to understand further the connection between the theory presented in \cite{doi:10.1073/pnas.2119288119,Mossa2023} and the older SPM \cite{PhysRevE.55.6917,10.1063/5.0158089}. We notice that the two frameworks start from completely different assumptions. The heterogeneous elasticity theory of \cite{doi:10.1073/pnas.2119288119,Mossa2023} follows a harmonic description, whose validity is arguable in the liquid phase, especially at large temperatures. On the other hand, the SPM starts from a anharmonic picture of the dynamics that involves additional low-energy modes. Also in this case, the nature and origin of these modes in liquids remains an open question. String-like excitations \cite{donati1998stringlike} can represent a candidate for these anharmonic excitations.
\section{Conclusions}\label{sec5}
\textit{What do instantaneous normal modes tell us about liquid dynamics?} Our work revolves around this question.

We have considered six different liquids, including Ar, Xe, Pb, Ga, N$_2$ and CS$_2$, in order to provide a (as much as possible) complete survey of INM properties and their relation to other standard probe for liquid dynamics, such as the velocity auto-correlation function, the self-diffusion constant and the dispersion of collective low-energy modes.

The highlights of our extensive analysis are collected below.
\begin{itemize}
    \item The linear scaling of the low-frequency INM DOS has been confirmed for all six liquids. This validates the hypothesis that, as far as the fraction of unstable mode is finite, the Debye scaling does not emerge in the INM spectrum, not even in the finite temperature solid phase. This is consistent with the idea that the linear scaling arises from overdamped unstable excitations that do not have propagating wave-like nature. Moreover, this linear scaling has been proven to coexist with a cusp-like singularity in the eigenvalue spectrum, showing that, contrary to recent claims in the literature \cite{doi:10.1073/pnas.2119288119}, the two features are not incompatible.
    \item The slope of the INM spectrum in the linear scaling regime $a(T)$ grows with temperature in simple liquids, following an Arrhenius-like exponential form, as previously demonstrated in \cite{jin2024temperature}. On the contrary, metallic liquids display an approximate constant slope, that results (at least in the range of $T$ investigated) independent of temperature.
    \item Both the fraction of unstable modes, but also the slope of the INM spectrum in the linear scaling regime $a(T)$, display a sharp jump at the first-order melting transition.
    \item The fraction of unstable modes, the momentum gap of shear collective waves (k-gap) and the self-diffusion constant display a very strong correlation in the whole liquid phase, confirming their role as ``probes of fluidity''.
    \item The temperature dependence of the fraction of unstable modes, but also the relation between the self-diffusion constant and the latter, do not appear to be universal in the different liquid systems considered.
    \item Against common expectations, crystals at finite temperature display a rather large number of unstable modes, that becomes negligible only at very low temperature. As a consequence, the scaling of the INM spectrum remains linear. At low temperature, a crossover scaling between a solid-like Debye scaling and a linear regime emerges in the INM spectrum, where the crossover frequency scale decreases with $T$. This is caused by the presence of non-diffusive unstable modes that survive in the solid phase. These modes are possibly a spurious effect of the naive INM computation and can be removed using several methods (\textit{e.g.}, \cite{10.1063/1.474968}). 
    \item In all liquids considered, the eigenvalue distribution displays a cusp-like singularity at $\lambda=0$. This feature can be explained both using heterogeneous-elasticity theory \cite{doi:10.1073/pnas.2119288119,Mossa2023}, but also using the soft potential model of liquids \cite{PhysRevE.55.6917,10.1063/5.0158089}. Moreover, the temperature dependence of the intermediate slopes and their difference as a function of temperature do not show the behavior previously reported in the literature for a simulated soft sphere liquid model \cite{doi:10.1073/pnas.2119288119,Mossa2023}. 
    \item As discussed in the main text, the linear in frequency DOS obtained from the INM analysis is compatible with the results from $k$-gap theory, suggesting an interesting connection between microscopic normal modes and collective excitations in liquids. Nevertheless, the INM analysis seems to underestimate the Debye quadratic contribution, specially close to the melting temperature. This could be an artifact of the INM approximation and in particular a result of the presence of unstable modes that do not vanish below the melting temperature. 
\end{itemize}

We conclude by noticing that many of these observations are yet not understood nor rationalized from a theoretical perspective, leaving an important hole in our understanding of liquid dynamics and interesting research directions for the near future. We hope that our work will play a motivating role in this direction.

\begin{acknowledgments}
We would like to thank T.~Keyes, J.~Douglas, A.~Zaccone, Y.~Wang, J.~Zhang, D.~Yu, C.~Stamper, J.~Moon, Y.~Yu, L.~Hong and H.~Xu for many discussions about INM and liquids. We also would like to thank  D. K. Belashchenko for instructions about the Gallium potential. We also thank T.~Keyes, J.~Douglas and J.~Moon for useful comments on a preliminary version of this manuscript. M.B. acknowledges the support of the Shanghai Municipal Science and Technology Major Project (Grant No.2019SHZDZX01) and the sponsorship from the Yangyang Development Fund. X. F. would like to thank the support provided by the National Natural Science Foundation of China (No. 52201016) and acknowledge the support of the State Key Laboratory of Clean and Efficient Turbomachinery Power Equipment for providing HPC resources. 
\end{acknowledgments}

\appendix
\subsection{Identification of the melting point for Argon}\label{ioio}
\begin{figure}[h]
    \centering
    \includegraphics[width=0.45\linewidth]{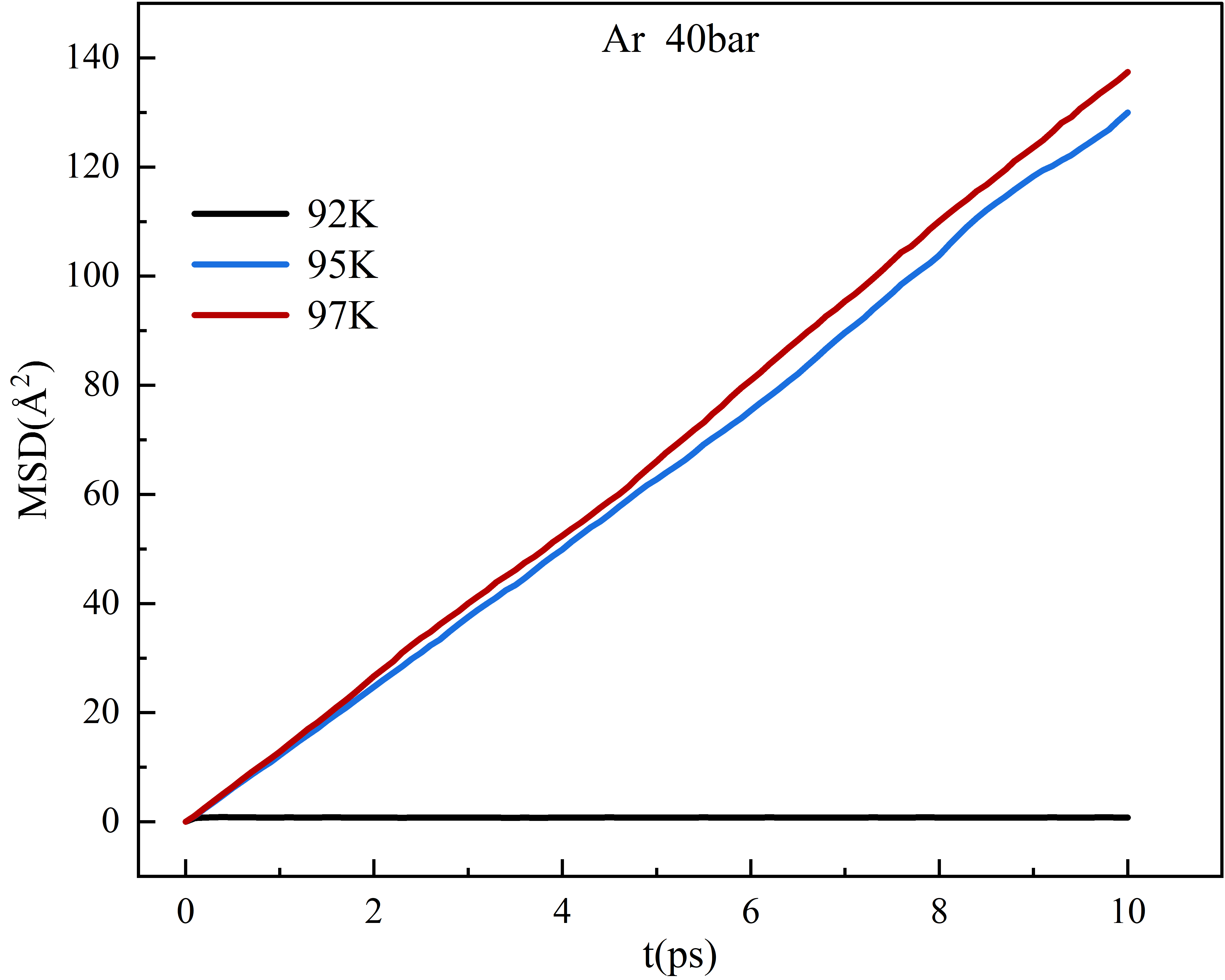}
    \caption{The mean square displacement of Argon at $40$ bar, measured at $92$ K, $95$ K and $97$ K.}
    \label{fig:msd}
\end{figure}

In fig.\ref{fig:msd}, we show the mean square displacement (MSD) as a function of time for Argon at $92$ K, $95$ K, and $97$ K at pressure of $40$ bar. At $92$ K, the behavior of the MSD differs significantly from that at $95$ K and $97$ K, demonstrating the absence of diffusion and confirming that the system is in the solid state. When the temperature rises to  $95$ K or above, the behavior of the MSD changes significantly, demonstrating a positive correlation with increasing temperature. This indicates a transition towards a liquid-like state. Using this heuristic argument, we can locate the position of the melting point within this temperature range.

\subsection{Additional data for Argon}
\begin{figure}[h]
    \centering
    \includegraphics[width=0.8\linewidth]{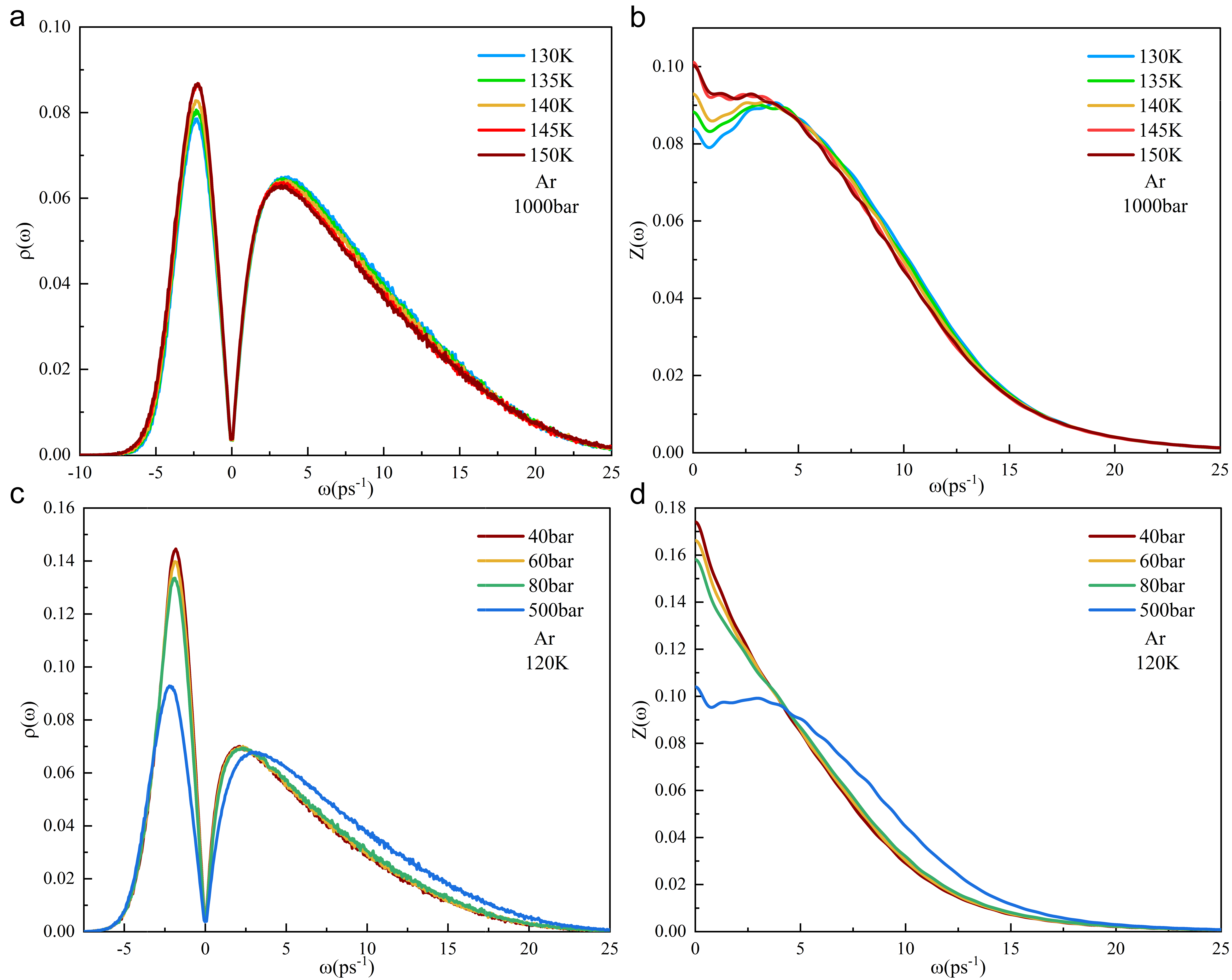}
    \caption{Panels \textbf{(a)} and \textbf{(b)} show the INM DOS $\rho(\omega)$ and the density of state function $S(\omega)$ for Argon at $1000$ bar as a function of temperature. Panels \textbf{(c)} and \textbf{(d)} show the same quantities at fixed temperature, $T=120$ K, as a function of the pressure.}
    \label{fig:si1}
\end{figure}

In the main text, we have shown the data for Argon only along the constant pressure line with $P=40$ bar. In Fig. \ref{fig:si1}, we report additional data at other pressures that confirm the findings discussed in the main text.

\subsection{Xenon, Nitrogen and more liquid systems}
\begin{figure}[h]
    \centering
    \includegraphics[width=0.8\linewidth]{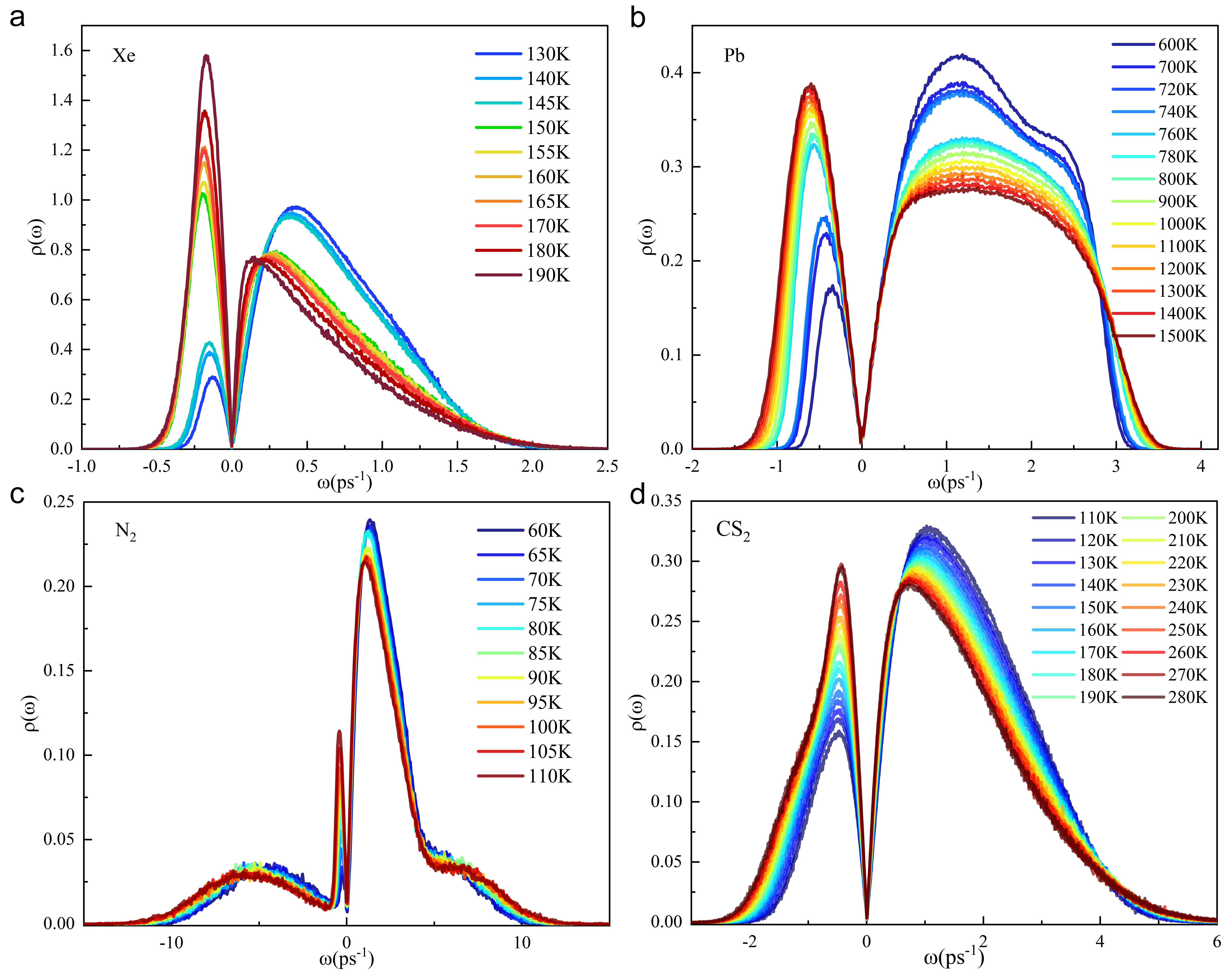}
    \caption{INM DOS $\rho(\omega)$ as a function of temperature for Xe \textbf{(a)}, Pb \textbf{(b)}, N$_2$ \textbf{(c)} and CS$_2$ \textbf{(d)}.}
    \label{fig:si2}
\end{figure}

\begin{figure}[h]
    \centering
    \includegraphics[width=0.8\linewidth]{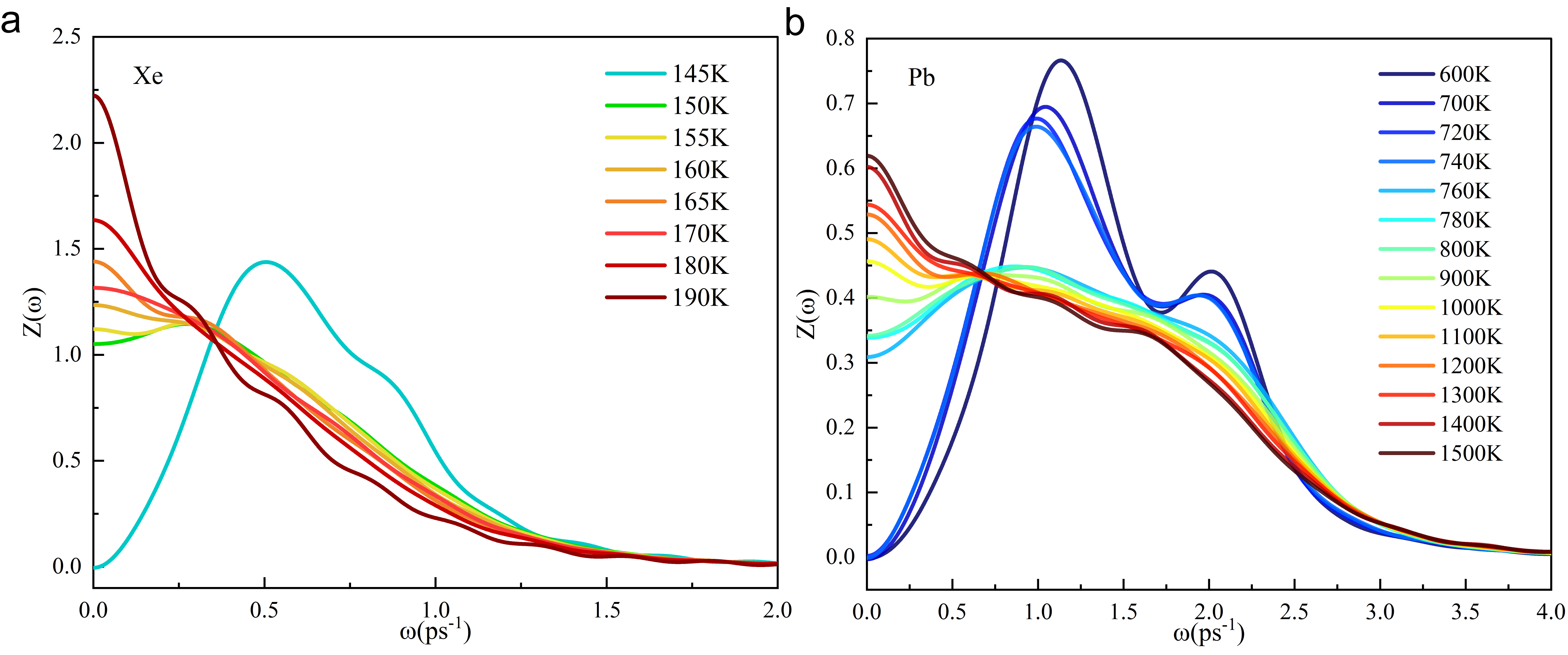}
    \caption{Density of state function $S(\omega)$ for Xe \textbf{(a)} and Pb \textbf{(b)} at different temperatures.}
    \label{fig:si3}
\end{figure}

Argon has been taken in the main text as a benchmark example for our analysis. Here, we report additional data for the other liquids discussed in this manuscript. Fig.~\ref{fig:si2} reports additional data for the INM DOS at different temperatures for Xe, Pb, N$_2$ and CS$_2$. Fig.~\ref{fig:si3} shows the density of state function at different temperature for Xe and Pb.

\subsection{Self-diffusion constant}\label{app:D}
\begin{figure}[h]
    \centering
    \includegraphics[width=\linewidth]{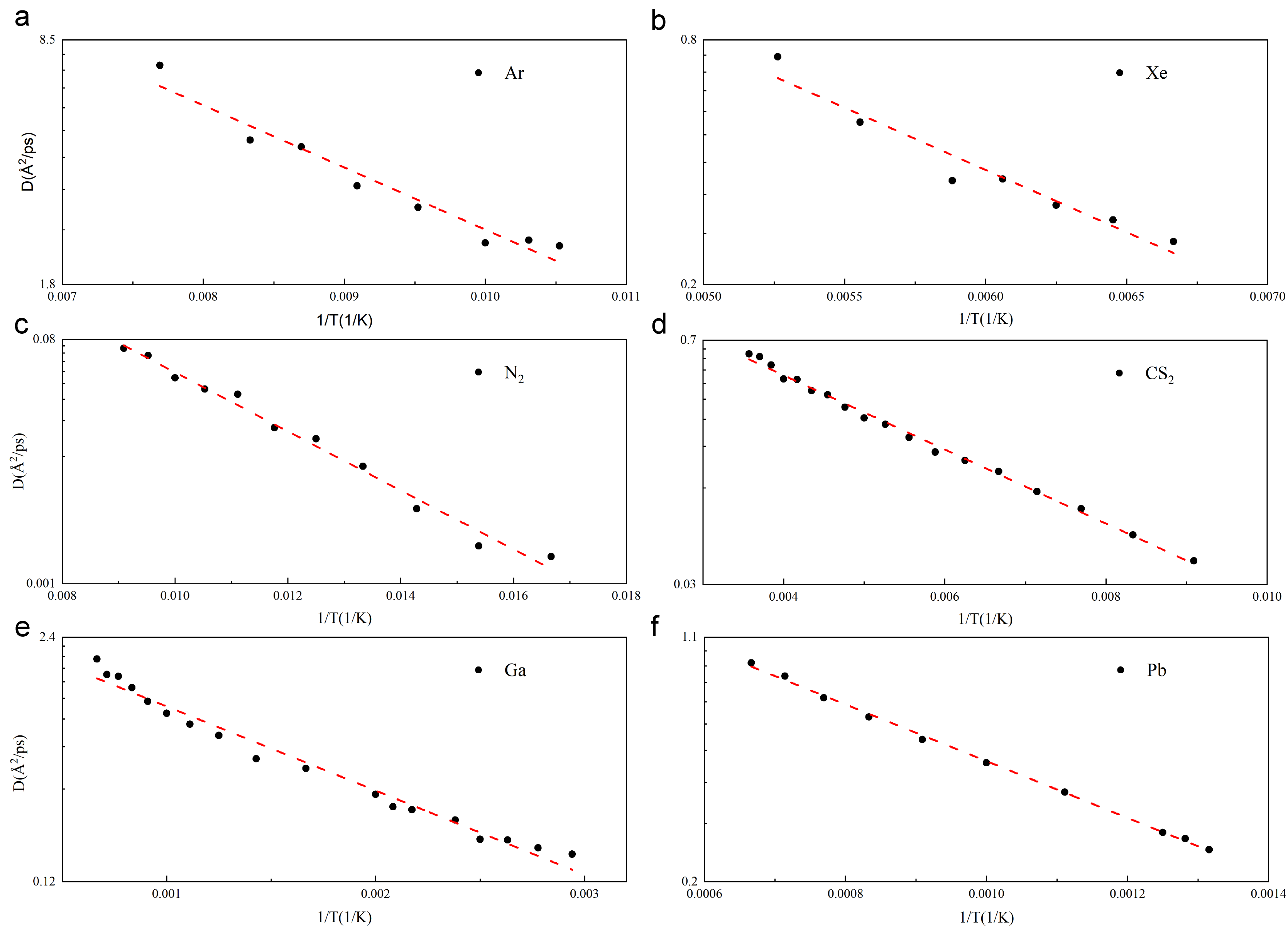}
    \caption{Arrhenius plot for the self-diffusion constant $D$ as a function of temperature. Different panels correspond to the six different liquids considered in this work: Ar \textbf{(a)}, Xe \textbf{(b)}, N$_2$ \textbf{(c)}, CS$_2$ \textbf{(d)}, Ga \textbf{(e)} and Pb \textbf{(f)}. The red dashed lines indicate the Arrhenius fit.}
    \label{fig:vvvvw2}
\end{figure}

In Fig. \ref{fig:vvvvw2} we show the Arrhenius plot for the self-diffusion constant of the six different liquid systems studied in our work. The corresponding activation energies are reported Eq.~\eqref{diffidiffi} in the main text.

\subsection{Slopes of the eigenvalue distribution $\nu(\lambda)$}\label{app:D}
\begin{figure}[h]
    \centering
    \includegraphics[width=0.8\linewidth]{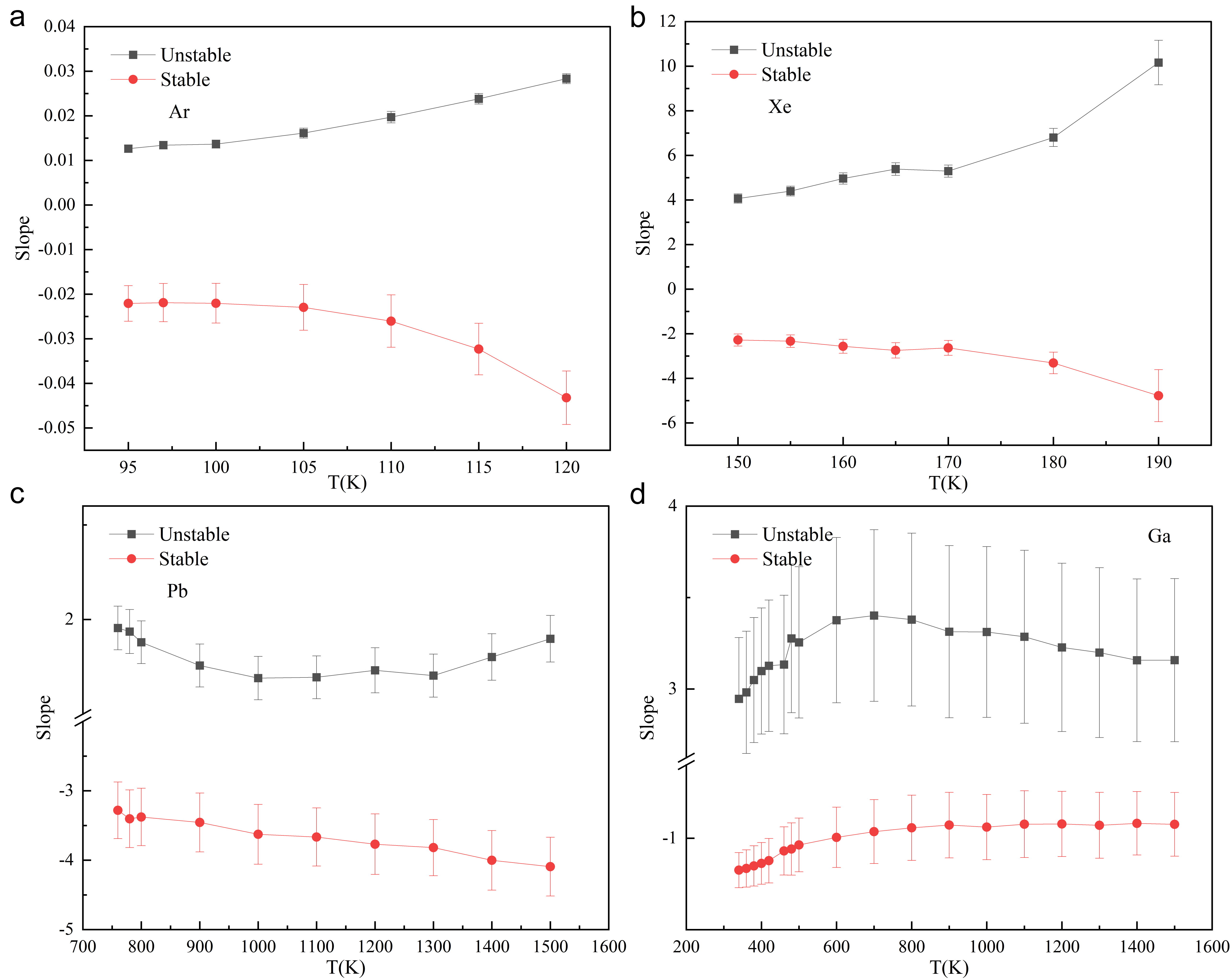}
    \caption{The ``slopes'' of the eigenvalue distribution $\upsilon(\lambda)$ as a function of temperature for Ar \textbf{(a)}, Xe \textbf{(b)}, Pb \textbf{(c)} and Ga \textbf{(d)}.}
    \label{fig:si5}
\end{figure}

In Fig. \ref{fig:si5}, we show the temperature behavior of the fitted stable and unstable slopes discussed in the main text for the other systems, including Ar, Xe, Pb and Ga.

\end{document}